\documentclass{aa}
\usepackage{graphicx}
\usepackage{natbib} \bibpunct{(}{)}{;}{a}{}{,}
\usepackage{color}
\usepackage{txfonts}

\def\changedA{}
\def\changedB{}
\def\changedC{}
\def\changedD{}
 
 \def\HI{\ion{H}{i}} \def\HII{\ion{H}{ii}} \def\HeI{\ion{He}{i}}
 \def\HeII{\ion{He}{ii}} \def\HeIII{\ion{He}{iii}} \def\CI{\ion{C}{i}}
 \def\CII{\ion{C}{ii}} \def\CIII{\ion{C}{iii}} \def\CIV{\ion{C}{iv}}
 \def\CV{\ion{C}{v}}  \def\NI{\ion{N}{i}} \def\NII{\ion{N}{ii}}
 \def\NIII{\ion{N}{iii}} \def\NIV{\ion{N}{iv}} \def\NV{\ion{N}{v}}
 \def\NVI{\ion{N}{vi}}  \def\OII{\ion{O}{ii}}
 \def\OIII{\ion{O}{iii}} \def\OIV{\ion{O}{iv}} \def\OV{\ion{O}{v}}
 \def\OVI{\ion{O}{vi}} \def\OVII{\ion{O}{vii}} 
 \def\SiIII{\ion{Si}{iii}} \def\SiIV{\ion{Si}{iv}} \def\SiV{\ion{Si}{v}}
 \def\FeI{\ion{Fe}{i}} \def\FeII{\ion{Fe}{ii}} \def\FeIII{\ion{Fe}{iii}}
 \def\FeIV{\ion{Fe}{iv}} \def\FeV{\ion{Fe}{v}} \def\FeVI{\ion{Fe}{vi}}
 \def\FeVII{\ion{Fe}{vii}} \def\FeVIII{\ion{Fe}{viii}} \def\FeIX{\ion{Fe}{ix}}
 \def\FeX{\ion{Fe}{x}}  
   
 \def\FeXVI{\ion{Fe}{xvi}} 
 \newcommand{\msunpyr}{\,M_\odot\,\mbox{yr}^{-1}}
\newcommand{\dint}{\,\mbox{d}} 

\newcommand{\kms}{\ifmmode{\,\mbox{km}\,\mbox{s}^{-1}}\else{km/s}\fi}
\newcommand{\msun}{\ifmmode M_{\odot} \else M$_{\odot}$\fi}
\newcommand{\rsun}{\ifmmode R_{\odot} \else R$_{\odot}$\fi}
\newcommand{\lsun}{\ifmmode L_{\odot} \else L$_{\odot}$\fi}
\newcommand{\zsun}{\ifmmode Z_{\odot} \else $Z_{\odot}$\fi}
\newcommand{\velo}{\ifmmode\varv\else$\varv$\fi}
\newcommand{\vinf}{\ifmmode\velo_\infty\else$\velo_\infty$\fi}
\begin{document} 
 
\title{Mass loss from late-type WN stars and its $Z$-dependence:}
\subtitle{very massive stars approaching the Eddington limit}

\titlerunning{Mass loss from late-type WN stars}
 
\author{G.\ Gr\"{a}fener \and W.-R.\ Hamann}
 
\institute{\centering Institut f\"ur Physik, Universit\"at Potsdam, Am Neuen
  Palais 10, D-14469 Potsdam, Germany}

\offprints{G.\ Gr\"afener\\
  \email{goetz@astro.physik.uni-potsdam.de}}
 
\date{Received ; Accepted}
 
\abstract{The mass loss from Wolf-Rayet (WR) stars is {{\changedD of
      fundamental importance for}} the {{\changedB final fate}} of massive
  stars and their chemical yields. Its $Z$-dependence is discussed in relation
  to the formation of {{\changedB {\changedC long-duration} Gamma Ray Bursts
      (GRBs) and the yields from early stellar generations.}}  However, the
  mechanism of formation of WR-type stellar winds is still under
  debate.}  {We present the first fully self-consistent atmosphere/wind models
  for late-type WN stars.  We investigate the mechanisms leading to their
  strong mass loss, and examine the dependence on stellar parameters, in
  particular on the metallicity $Z$.}  {We perform a systematic parameter
  study of the mass loss from WNL stars, utilizing a new generation of
  hydrodynamic non-LTE model atmospheres.  The models include a
  self-consistent treatment of the wind hydrodynamics, and take Fe-group
  line-blanketing and clumping {\changedA into account}.  They thus allow a
  realistic modelling of the expanding atmospheres of WR\,stars.  The results
  are verified by comparison with observed WNL spectra.}  {We identify WNL
  stars as very massive stars close to the Eddington limit, potentially still
  in the phase of central H-burning.  Due to their high $L/M$ ratios, these
  stars develop optically thick, radiatively driven winds. {{\changedB These
      winds show qualitatively different properties than the thin winds of OB
      stars.}}  The resultant mass loss {{\changedB depends strongly}} on $Z$,
  but also on the Eddington factor $\Gamma_{\rm e}$, and the stellar
  temperature $T_\star$. We combine our results in a parametrized mass loss
  recipe for WNL stars.}  {According to our present model computations, stars
  close to the Eddington limit tend to form strong WR-type winds, even at very
  low $Z$.  Our models thus predict an efficient mass loss mechanism for low
  metallicity stars. For extremely metal-poor stars, we find that the
  self-enrichment with primary nitrogen can drive WR-type mass loss.  These
  first WN\,stars might play an important role in the enrichment of the early
  ISM with freshly produced nitrogen.}  {} 
\maketitle

\section{Introduction} 
\label{sec:intro} 

The strong mass loss from Wolf-Rayet (WR) stars fundamentally affects the
evolution, the final fate, and the chemical yields of the most massive stars.
Its dependence on metallicity ($Z$) received renewed interest because of the
possible importance for early generations of massive stars and their chemical
yields \citep[see][]{mey1:06,chi1:06}. Moreover, within the collapsar model
\citep{woo1:93,fad1:99}, fast rotating WR stars are now broadly accepted as
the progenitors of {\changedC long-duration} Gamma Ray Bursts (GRBs). The formation
of GRB progenitors with the demanded properties could be explained by
lowered WR mass loss rates {\changedA at low metallicities}
\citep{yoo1:05,woo1:06}.

Despite their importance, the mechanism of formation of WR-type
stellar winds is still under debate.  The high WR luminosities and surface
temperatures suggest that WR winds are driven by radiation.  The
observed wind performance numbers $\eta = \dot{M}\varv_\infty/(L_\star/c)$, on
the other hand, {\changedC appear to} challenge this hypothesis.  Commonly observed
values in the range $\eta = 1$--10 imply a mechanical wind momentum much in
excess of the momentum of the radiation field. If WR winds are
driven by radiation, photons must be used more than once. {\changedC This can
  either be achieved by multiple line scattering, or by successive
  re-distribution and absorption of thermalized photons.}

The question of the driving mechanism of WR mass loss is strongly related to
its dependence on metallicity ($Z$). If WR winds are driven by radiation, one
would expect a similar $Z$-dependence as for the radiatively driven winds of
OB stars.  Such a relation has recently been claimed by \citet{vin1:05} for
the winds of {late} WR subtypes (WNL and WCL).
Because of the absence of observations over a broad metallicity range,
computational wind models are needed to address these questions.  However, as
previously mentioned, WR winds are far from being fully understood.
Consequently only a few models are available, partly based on controversial
model assumptions.

The first models for WR winds were constructed by \citet{luc1:93} and
\citet{spr1:94}, using Monte-Carlo techniques. These authors could show that
multiple line scattering can account for the high observed wind performance
numbers (up to $\eta\,$$\approx\,$$10$) {if} enough line opacities are
present. However, the models could not pove whether WR winds are entirely
driven by radiation, because they rely on an adopted velocity structure,
without solving the wind hydrodynamics.  In particular the models failed to
reproduce the wind acceleration in deep atmospheric layers.  {\changedA The WNL
  models by \citet{dek1:97} and} the $Z$-dependent models by \citet{vin1:05}
are based on a similar modelling technique where the ionization conditions in
the extended WR atmospheres are inferred from model atmosphere grids.  Also
these models cannot explain the wind driving in deep layers.

A completely different approach was followed by \citet{nug1:02} and
\citet{lam1:02}. By means of a critical-point analysis these authors could
show that the observed WR mass loss rates are in agreement with the assumption
of optically thick, radiatively driven winds.  In this approach it is assumed
that the sonic point is located in very deep atmospheric layers where the
diffusion limit is valid (note that under this assumption the sonic point
becomes the critical point of the wind flow). \citeauthor{nug1:02} found that
the sonic point must be located in specific temperature regimes where the
Rosseland mean opacity is increasing due to the well-known Fe opacity peaks
\citep[cf.][]{igl1:96}. In this picture, late-type WR winds are {\changedA
  initiated by} the ``cool Fe-peak'' (for sonic point temperatures of $T_{\rm
  s} = 40$--70\,kK), whereas the winds of early subtypes are supported by the
``hot Fe-peak'' ($T_{\rm s} > 160$\,kK).

Our own approach follows a computationally much more expensive path, where
{\changedA the full wind is modeled in a self-consistent manner. For this
  purpose a hydrodynamic solution scheme is incorporated into the Potsdam
  Wolf-Rayet (PoWR) atmosphere models. Within these models a detailed non-LTE
  radiative transfer is performed, which takes Fe-group line-blanketing and
  wind clumping into account.  They are thus able to describe the
  conditions in} the WR wind, including the optically thick part, in a
realistic manner (see Sect.\,\ref{sec:models} for a detailed description of
the PoWR models).
Using these models we could show that it is in principle possible to
drive a strong WR wind by radiation pressure alone \citep{gra1:05}.  In
accordance with \citet{nug1:02}, we found that the inner wind regions of
early-type WC stars are supported by line opacities from the hot Fe-peak
(ionization stages \FeIX--\FeXVI), and that the observed mass loss rates can
only be reached for extremely high critical-point temperatures
($\approx$\,$200$\,kK). Moreover, we could demonstrate the importance of
clumping in the wind dynamics, an effect which has not been taken into
account in any previous wind calculation.

In the present work we focus on late-type WN\,stars (WNL\,stars, spectral
subtypes WN\,6-9). Compared to early subtypes, these stars have rather low
effective temperatures ($30$--$60\,$kK) in a similar range as, e.g., typical O
supergiants.  \citet{ham1:06} recently found that these stars form a distinct
group within the galactic WN {\changedA population, with exceptionally high
  luminosities (above $\sim$\,$10^{5.9}\,L_\odot$). This result is in line
  with previous detections of very high luminosities for H-rich WNL subtypes
  in the LMC \citep{dek1:97,cro1:98}, and the galactic center \citep{naj1:04}.
  Here} we investigate the question of why these stars show enhanced, WR-type
mass loss, and present a grid of WNL wind models for different metallicities.
Moreover, we examine the possibility of strong WR-type mass loss for the
earliest generations of metal-poor massive stars, {\changedC as a result of}
self-enrichment with primary nitrogen \citep[see][]{mey1:06}.

{\changedA In Sect.\,\ref{sec:atmos} we give a short overview of the applied
  modelling technique and the {\changedC adopted} atomic data.  In
  Sect.\,\ref{sec:models} we present the results of our model computations and
  compare them with observations.  In Sect.\,\ref{sec:wdrv} we discuss the
  specific properties of WR-type winds, and relate them to previous mass loss
  studies. Finally, we discuss the most important implications of our results
  in Sect.\,\ref{sec:discussion}, and summarize the main conclusions in
  Sect.\,\ref{sec:conclusions}.}

\section{Atmosphere models}
\label{sec:atmos}

The Potsdam Wolf-Rayet (PoWR) model atmosphere code is a state-of-the-art code
for expanding stellar atmospheres, which incorporates the treatment of
line-blanketing due to millions of Fe-group transitions in non-LTE \citep[for
details of the numeric treatment
see][]{koe1:92,ham1:92,leu1:94,koe1:95,leu1:96,ham1:98,koe1:02,
  gra1:02,ham1:03}. Within our group this code has been extensively applied
to the quantitative analysis of WR\,stars
\citep{gra1:00,gra1:02,ham3:03,sta1:04,ham1:06,bar2:06}. Recently the
code was extended by a self-consistent solution of the hydrodynamic equations
\citep{gra1:05}, where the atmosphere structure ($\rho(r)$ and $\varv(r)$) is
determined consistently with the radiative acceleration as obtained from the
non-LTE radiation transport.

{\changedC Here} we give {\changedA an overview of the applied modelling
  technique.} In Sect.\,\ref{sec:numerics} we describe the applied numerical
methods, in Sect.\,\ref{sec:mpar} we review the relevant model parameters, and
in Sect.\,\ref{sec:atoms} we summarize the atomic data utilized for the
calculations.

\subsection{Numerical methods}
\label{sec:numerics} 

The model code computes the radiation field, the atomic {\changedA level}
populations, the temperature structure, the density structure, and the
velocity field for a stationary, spherically expanding stellar atmosphere. The
complete solution comprises four parts which are iterated until consistency is
obtained: the radiation transport, the equations of statistical equilibrium,
the energy equation, and the hydrodynamic equations.

The radiation transport for a spherically expanding atmosphere is formulated
in the co-moving frame of reference (CMF), neglecting aberration and advection
terms \citep[see][]{mih1:76}.  For a fast solution and a consistent treatment
of electron scattering we employ the method of variable Eddington factors
\citep{aue1:70}. This means that the moment equations are solved to obtain the
angle-averaged radiation field, and the numerically expensive ray-by-ray
transfer is only calculated from time to time \citep{koe1:02,gra1:02}.  The
fast numerical solution allows a detailed treatment of millions of spectral
lines on a fine frequency grid.

The atomic populations and the electron density are determined from the
equations of statistical equilibrium.  This system of equations is solved in
line with the radiation transport by application of the ALI formalism
\citep[accelerated lambda iteration, see][]{ham2:85,ham1:86}.  Complex model
atoms of He, C, {\changedA N,} O, Si, and the Fe-group are taken into account.
For the inclusion of millions of iron-group transitions we take advantage of
the super-level concept following \citet{and1:89}.  {\changedC The details of the
  numerical implementation are described by \citet{gra1:02}}.  Density
inhomogeneities are taken into account {\changedC in the limit of} small-scale
clumps, {\changedA or micro-clumping {\changedC\citep[see][and
    Sects.\,\ref{sec:clump} and \ref{sec:dcl} in the present work]{ham1:98}}.
  The clumping factor $D$ denotes the density enhancement within a clump, or
  the inverse of the volume filling factor $f_V$.}

The temperature structure is obtained from the assumption of radiative
equilibrium.  In the present work this constraint equation is decoupled from
the equations of statistical equilibrium. It is solved by a temperature
correction procedure which is based on the Uns\"old-Lucy method
\citep{uns1:55,luc1:64}, and has been generalized for application in non-LTE
models with spherical expansion \citep{ham1:03}.

The density- and velocity structure of the expanding atmosphere is obtained
from the hydrodynamic equations for a stationary radial flow, accelerated by
radiation pressure \citep[see][]{gra1:05}. The radiative acceleration is
obtained by direct integration within the CMF radiation transport
\begin{equation}
\label{eq:arad}
a_{\rm rad} = \frac{1}{\rho}\,\frac{4\pi}{c} \int_0^{\infty} \kappa_\nu
H_\nu\,\dint \nu,
\end{equation}
where the Eddington flux $H_\nu$ and the opacity $\kappa_\nu$ are both
computed on a fine frequency grid.

Due to Doppler shifts, the radiative acceleration on spectral lines reacts to
changes of the velocity gradient $\velo'$. Because of its influence on the
dynamic properties of the gas flow, this dependence must be included in the
hydrodynamic solution. In our computations we perform an additional radiative
transfer calculation in advance of each hydro-iteration, where the velocity
distribution $\velo(r)$ is changed.  In this way we determine the fraction of
$a_{\rm rad}$ that responds linearly to small variations of $\velo'$.
{\changedA This value corresponds to the effective contribution of optically
  thick spectral lines ($a_{\rm thick}$) to the total radiative acceleration
  \citep[see, e.g.,][]{pul1:00}.  Dividing $a_{\rm thick}$ by the total
  acceleration from spectral lines ($a_{\rm lines}$) gives the force
  multiplier parameter $\alpha$, which} has a similar meaning as the parameter
$\alpha$ in the standard theory of line driven winds by
\citet[][CAK]{cas1:75}. {\changedA Note, however, that our $\alpha$ is an {\em
    effective} value which includes complex effects like line overlaps and
  changes in the wind ionization, while the CAK $\alpha$ is deduced from
  atomic line statistics.}

\subsection{Model parameters} 
\label{sec:mpar}

{\changedA Our hydrodynamic} model atmospheres are characterized by the mass,
the luminosity, the radius, and the chemical composition of the stellar core.
Moreover, the clumping factor must be prescribed {\changedA throughout the
  wind.}
Given these parameters, the density and velocity structure of the atmosphere
follow from the hydrodynamic solution. The basic model parameters are: the
stellar core radius $R_\star$ at Rosseland optical depth $\tau_\mathrm{R} =
20$, the stellar temperature $T_\star$ (related to the luminosity $L_\star$
and the core radius $R_\star$ via the Stefan Boltzmann law), the stellar mass
$M_\star$, the chemical composition, and the clumping factor $D$ which is
prescribed as a function of $\tau_\mathrm{R}$. Note that we define
$\tau_\mathrm{R}$ in this context as the `LTE-continuum' Rosseland mean
opacity.

For the line broadening we assume Doppler profiles with a broadening velocity
of $\varv_{\rm D} = 100\kms$ throughout this work.  In this way the absorption
troughs of P-Cygni type line profiles are successfully reproduced. Moreover,
we are presently limited to relatively high {\changedA broadening velocities}
because the frequency spacing in our models scales with $\varv_{\rm D}$, and
the computation time thus scales with $1/\varv_{\rm D}$.

Because of its effect on the radiative acceleration, the detailed prescription
of the clumping factor {\changedA$D(r)$} has an influence on the obtained
results.  {\changedA Unfortunately}, this value can only be constrained very
roughly from the {\changedA observed} electron scattering wings of strong
emission lines \citep[see][]{ham1:98}.  For the {\changedA outer part of the
  wind} we thus adopt a constant value of {\changedA $D_{\rm max}=10$}, which
gave satisfactory results in previous {\changedC analyses of
  similar stars \citep[e.g.,][]{her1:01,mar1:08}}. For the radial dependence
in the {\changedA deeper wind layers} we rely on previous results for WR\,111
\citep{gra1:05} where a $\tau$-dependent formulation reproduces
high ionization stages very well, which are formed deep inside the atmosphere.
In this formulation $D$ continuously increases from a value of {\changedA
  $D_{\rm min}=1$} {\changedA at the wind base} (for $\tau_\mathrm{R} > 0.7$)
to {\changedA $D_{\rm max}=10$} in the outer layers (for $\tau_\mathrm{R} <
0.35$).

The freedom in the prescription of $D(r)$ is not satisfactory. However,
clumping is clearly detected in WR\,winds and should be included in the
models. As long as only optically thin regions above the critical point of the
equation of motion are affected by clumping, its influence on the obtained
mass loss rates is {\changedA relatively} small (see Sect.\,\ref{sec:clump}).
Nevertheless, we expect a strong influence on the outer wind structure and in
particular on the terminal wind velocities.

\subsection{Atomic data}
\label{sec:atoms} 

\begin{table}[t]
\begin{center}
\begin{tabular}{lllllll}
\hline \hline 
\rule{0cm}{2.2ex}Ion & Levels & Ion & Levels & Ion & Super-  & Sub- \\
                     &        &     &        &     & levels  & levels\\
\hline
\rule{0cm}{2.2ex}\HI&     10 &\NIV&    44 &\FeI&     1 & 21 \\    
\HII&     1 &\NV&     17 &\FeII&   12 & 17170 \\ 
\HeI&    17 &\NVI&     1 &\FeIII&  16 & 14188 \\ 
\HeII&   16 &\OII&     3 &\FeIV&   18 & 30122 \\ 
\HeIII&   1 &\OIII&    9 &\FeV&    19 & 19804 \\ 
\CI&      1 &\OIV&     8 &\FeVI&   18 & 15155 \\ 
\CII&     3 &\OV&     12 &\FeVII&  16 & 11867 \\ 
\CIII&   12 &\OVI&     9 &\FeVIII& 17 & 8669  \\ 
\CIV&     9 &\OVII&    1 &\FeIX&   19 & 12366 \\ 
\CV&      1 &\SiIII&  10 &\FeX&     1 & 1 \\     
\NI&      3 &\SiIV&    7 &&&\\
\NII&    38 &\SiV&     1 &&&\\
\NIII&   40 &&&&&\\
\hline
\end{tabular} 
\end{center}
\caption{Summary of the model atom. Fe-group ions (Fe)
  are described by a relatively small number of super-levels, each 
  representing a large number of true atomic energy levels (sub-levels).
  {\changedA The relative abundances of the Fe-group elements are listed in 
    \citet{gra1:02}, Table~2.}
} 
\label{tab:modelatom}
\end{table}

The chemical composition of the model atmospheres is given by {\changedC the} mass
fractions $X_{\rm He}$, $X_{\rm C}$, $X_{\rm N}$, $X_{\rm O}$, $X_{\rm Si}$,
and $X_{\rm Fe}$ of helium, carbon, nitrogen, oxygen, silicon, and iron-group
elements.  Compared to our previous models for early-type WC\,stars
\citep{gra1:02,sta1:04,gra1:05} we have reduced the model atoms significantly.
This is possible because of the lower effective temperatures, and the specific
chemical composition of WN\,stars (mainly H, He, and N).  In particular, we
use rather small atomic models for carbon and oxygen, and a reduced number of
ionization stages for the iron group {\changedC (\FeI--{\sc x})}.  In this way we manage to
combine the most relevant atomic species in a model atom with 410 super
levels.  This model atom still allows moderate computing times of the order of
1--3 days {\changedA for a hydrodynamic} model, i.e., it permits
pa\-ra\-me\-ter studies with a large number of grid models.

The model atoms are summarized in Table\,\ref{tab:modelatom}.  The atomic data
are compiled from the following sources.  Oscillator strengths for CNO are
taken from the Opacity Project
\citep{hum1:88,sea1:87,sea1:92,cun1:92,sea1:94,the1:95}, level energies from
Kurucz's CD-ROM No.\,23 \citep{kur1:95}, and ionization cross sections from
\citet{nah1:97} and \citet{nah1:99}. Fe-group data are taken from Kurucz
CD-ROM No.\,20--22 \citep[see][]{kur1:91,kur1:02}.

\section{Model calculations}
\label{sec:models}

In this section we present a systematic parameter study of the mass loss from
late-type WN\,stars. The first question concerning these objects is {\em why}
they become WR\,stars, i.e., why they show enhanced mass loss.
We address this question in Sect.\,\ref{sec:test},
where we investigate the dependence of WNL mass loss on stellar parameters,
and compare our results with observed properties of the galactic WNL sample.
In Sect.\,\ref{sec:wr22} we perform a more detailed comparison with WR\,22, a
spectroscopic binary {\changedA for which} the mass of the WR component is
roughly known.  {\changedA Finally, we investigate the $Z$-dependence of WNL
  star mass loss in Sect.\,\ref{sec:zdep}, and combine our results in the form
  of a mass loss recipe in Sect.\,\ref{sec:recipe}.}

\subsection{WNL\,stars at solar metallicity}
\label{sec:test}

\begin{figure}[t!]
\parbox[b]{0.49\textwidth}{\includegraphics[scale=0.4]{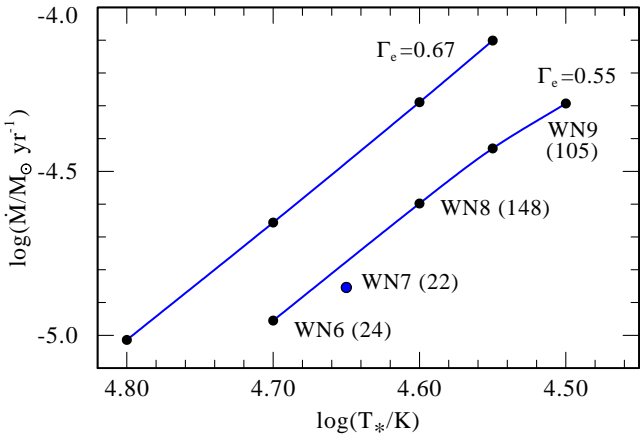}}
\caption{Wind models for galactic WNL stars:
  mass loss rates for different stellar temperatures $T_\star$ and Eddington
  factors $\Gamma_{\rm e}$. The corresponding spectral subtypes are indicated
  together with WR numbers of specific galactic objects \citep[according to][
  in brackets]{huc1:01}, which show a good agreement with the synthetic
  line spectra.  Note that the models are computed for a fixed stellar
  luminosity of $10^{6.3}\,L_\odot$.  The WN\,7 model (WR\,22) is slightly
  offset from the standard grid models because it is calculated with an
  enhanced hydrogen abundance (see Sect.\ref{sec:wr22}).
  \label{fig:mdot-wnl}
}
\end{figure}

\begin{figure}[t!]
\parbox[b]{0.49\textwidth}{\includegraphics[scale=0.4]{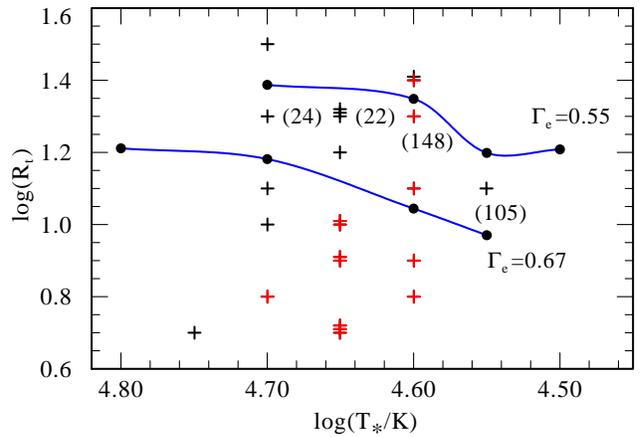}}
\caption{Wind models for galactic WNL stars: wind densities compared
  to observed values according to \citet{ham1:06}.  In the $R_{\rm
    t}$--$T_\star$ plane, our models (solid curves) reproduce the {\em upper}
  part of the observed WNL sample (crosses), corresponding to WNL stars with
  {\em low} wind densities (see text for the definition of $R_{\rm t}$). The
  objects belonging to the enigmatic class of WN\,8 spectral subtypes are
  indicated in red.
  \label{fig:rtts}
}
\end{figure}

\begin{figure}[t!]
\parbox[b]{0.49\textwidth}{\includegraphics[scale=0.4]{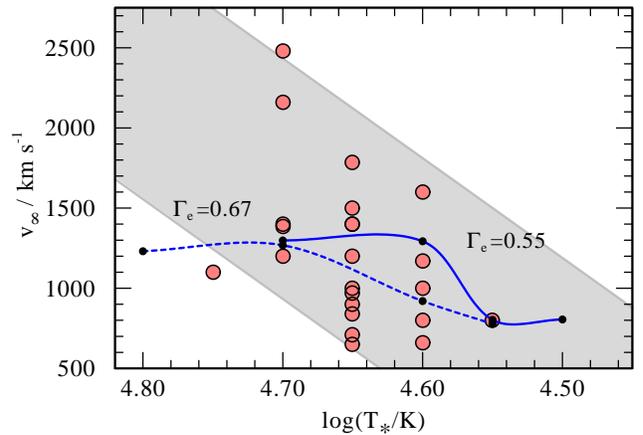}}
\caption{Wind models for galactic WNL stars: terminal wind velocities compared
  to observed values {\changedA taken from \citet{ham1:06}, and the observed
    trend with spectral subtype \citep[from][]{nie1:02}}. Our models (solid
  curve for $\Gamma_{\rm e}=0.55$ and dashed curve for $\Gamma_{\rm e}=0.67$)
  reproduce the observed trend of increasing $\varv_\infty$ with increasing
  $T_\star$, but they tend to underestimate the high values observed for the
  hottest objects.
  \label{fig:vinf-wnl}
}
\end{figure}

 \begin{table*}[] 
{\changedA
  \begin{center}
  \begin{tabular}{lllllllll} 
    \hline \hline 
    \rule{0cm}{2.2ex}Models & Grid\,1: $T_\star$, $\Gamma_{\rm e}$ & WR\,22     &
    Grid\,2: $Z$, $\Gamma_{\rm e}$ & Grid\,3: $X_{\rm Fe}$\,$=$\,$0$  & Grid\,4:
    $X_{\rm H}$, $T_\star$ & Grid\,5: $L_\star$, $T_\star$ & VdK05 \\
    \hline
    \rule{0cm}{2.2ex}$L_\star /  L_\odot $    & $10^{6.3}$ & $10^{6.3}$  &
    $10^{6.3}$  & $10^{6.3}$ & $10^{6.3}$ & $10^{5.1...6.6}$ & $10^{5.6}$ \\
    $R_\star / R_\odot$     & 47.2\,...\,11.9      & 23.7      & 23.7          & 23.7 & 18.8\,...\,47.2 & 4.7\,...\,66.7 & 13.5\\ 
    $T_\star / \mathrm{kK}$ & 31.6\,...\,63.1     & 44.7      & 44.7          & 44.7  &  31.6\,...\,50.1 & 31.6\,...\,50.1 & 40.0 \\ 
    $M_\star / M_\odot$     & 67, 55          & 78.1      & 101\,...\,48.4    & 59.3\,...\,49.1 & 55.8\,...\,100.4 & 4.9\,...\,155.9 & 20.0 \\
    $\Gamma_{\rm e}$        & 0.55, 0.67     & 0.55       & 0.44\,...\,0.89  & 0.73\,...\,0.88 & 0.55 & 0.55 & 0.37\\
    {\changedA $D_{\rm max}$} & 10              & 10       & 10          & 10 & 10 & 10 & 10\\
    \hline 
    \rule{0cm}{2.2ex}$X_\mathrm{H}$  & 0.2 & 0.4 & 0.4  & 0.4 &  0.0\,...\,0.8 & 0.4 & 0.15\\
    $X_\mathrm{He}$ & 0.78           & 0.58 & 0.58  & 0.58 & 0.18\,...\,0.98 & 0.58 & 0.83\\
    $X_\mathrm{C}$  & $4.0\,10^{-4}$ & $4.0\,10^{-4}$ & $4.0\,10^{-4} \cdot z$ & $4.0\,10^{-4}$ & $4.0\,10^{-4}$ & $4.0\,10^{-4}$ & $4.0\,10^{-4}$\\
    $X_\mathrm{N}$  & $1.4\,10^{-2}$ & $1.4\,10^{-2}$ & $1.4\,10^{-2} \cdot z$ & $1.4\,10^{-2}$ & $1.4\,10^{-2}$ & $1.4\,10^{-2}$ & $1.4\,10^{-2}$\\
    $X_\mathrm{O}$  & $1.0\,10^{-3}$ & $1.0\,10^{-3}$ & $1.0\,10^{-3} \cdot z$ & $1.0\,10^{-3}$ & $1.0\,10^{-3}$ & $1.0\,10^{-3}$ & $1.0\,10^{-3}$\\
    $X_\mathrm{Si}$ & $8.0\,10^{-4}$ & $8.0\,10^{-4}$ & $8.0\,10^{-4} \cdot z$ & 0 & $8.0\,10^{-4}$ & $8.0\,10^{-4}$ & $8.0\,10^{-4}$\\
    $X_\mathrm{Fe}$ & $1.6\,10^{-3}$ & $1.6\,10^{-3}$ & $1.6\,10^{-3} \cdot z$ & 0 & $1.6\,10^{-3}$ & $1.6\,10^{-3}$ & $1.6\,10^{-3}$\\
    \hline
    \rule{0cm}{2.2ex}$\dot{M} / \msunpyr$ & $10^{-(4.1\,...\,5.0)}$ & $10^{-4.85}$
    & $10^{-(4.6\,...\,5.4)}$ & $10^{-(4.7\,...\,5.5)}$  & $10^{-(4.2\,...\,5.2)}$ & $10^{-(4.2\,...\,5.5)}$  & $10^{-(5.41)}$ \\
    $\vinf / \kms$                        & 778\,...\,1298  & 974  & 240\,...\,1903  & 388\,...\,478 & 646\,...\,1396 & 620\,...\,1567 & 941\\ 
    $R_\mathrm{t} / R_\odot$              & 9.6\,...\,20.6  & 21.7 & 9.2\,...\,53.7  & 9.1\,...\,37.7 & 12.5\,...\,39.7 & 8.02\,...\,30.7 & 28.2\\ 
    $\eta$                                & 0.28\,...\,1.51 & 0.34 & 0.04\,...\,1.15  & 0.04\,...\,0.21 & 0.20\,...\,1.12 & 0.20\,...\,1.72 & 0.44\\ 
    \hline 
  \end{tabular}
  \end{center}
  \caption{Model parameters for our different (grid) computations.
    The first column (Grid\,1) denotes the models from
    Sect.\,\ref{sec:test} for solar $Z$, where $T_\star$ and $\Gamma_{\rm e}$ are varied.
    In the second column the parameters for the WR\,22 model from
    Sect.\,\ref{sec:wr22} are given. The third column (Grid\,2) indicates our
    $Z$-dependent models from Sect.\,\ref{sec:zdep}, followed by 
    the computations for extremely metal-poor stars with zero Fe
    abundance (Grid\,3). The {\changedC next} two columns
    belong to Sect.\,\ref{sec:recipe} where the dependence on
    $X_{\rm H}$ (Grid\,4) and $L_\star$ (Grid\,5) are investigated.
    The last column (VdK05) indicates a model from Sect.\,\ref{sec:comp}
    for comparison with the WNL model from \citet{vin1:05} for solar $Z$.
    Note that for the $Z$-dependent models $z = Z/Z_\odot$ indicates the
    relative mass fraction of metals with respect
    to the solar value. The 
    luminosity $L_\star$, the radius $R_\star$, the temperature $T_\star$,
    and the mass $M_\star$ of the stellar core
    (with the corresponding Eddington factor $\Gamma_{\rm e}$),
    the assumed maximum wind clumping factor {\changedA $D_{\rm max}$}, the surface
    mass fractions
    $X_{\rm He}$, $X_{\rm C}$, $X_{\rm H}$, $X_{\rm O}$, $X_{\rm Si}$, and
    $X_{\rm Fe}$, and the resultant mass loss rates $\dot{M}$,
    transformed radii $R_{\rm t}$, terminal wind velocities $\vinf$,
    and wind efficiencies $\eta = \dot M \varv_\infty/(L_\star/c)$ are given.}
  \label{tab:wrpar} 
}
\end{table*}

A large part of the galactic WN population has been re-analyzed by
\citet{ham1:06}, {\changedA based on a publicly accessible grid of
  line-blanketed atmosphere models \citep[][{\tt
    http://www.astro.physik.uni-potsdam.de}]{ham1:04}.}  In this study the WN
parameters have been significantly revised \citep[{\changedA e.g.,} compared
to][]{ham2:98}, chiefly due to the influence of Fe-group line-blanketing, and
the consideration of 2MASS IR photometry.  In particular, the galactic WN
stars are now found to form two distinct groups which are divided by their
luminosities.  The first group consists of a mixture of early- to intermediate
{\changedA WN subtypes} with luminosities below {\changedA $10^{5.9}\,L_\odot$}.
The second group, formed by the WNL stars (WN\,6--9), shows luminosities above
this value.  While the earlier subtypes are dominated by pure He-stars, WNL
stars {\changedA tend to} show a significant amount of hydrogen at their
surface (up to $X_{\rm H}=0.55$).  They are located to the right of the
zero-age main-sequence in a temperature range of $T_\star = 35$--55\,kK, and
thus might still be in the phase of central H-burning.
In the present work we concentrate on this class of extremely luminous evolved
stars, which are in many cases among the most luminous objects in {\changedA
  young} stellar clusters.

Our main goal now is to show why these stars become WR\,stars, or in other
words, why they show enhanced mass loss. WNL stars have similar
effective temperatures and surface compositions as typical O-supergiants
\citep[see e.g.][]{cro2:02} but they show much stronger wind emission lines.
To investigate their mass loss properties we have prepared a series of
hydrodynamic PoWR models with typical WNL parameters \citep[referring
to][]{ham1:06}.  We adopt a high luminosity of $L_\star=10^{6.3}\,L_\odot$,
effective core temperatures in the range $T_\star=33$--55\,kK, a clumping
factor of {\changedA $D_{\rm max}=10$}, and typical WN surface abundances with
$X_{\rm H}=0.2$, $X_{\rm He}=0.78$, and $X_{\rm N}=0.014$ (for more details
see Table\,\ref{tab:wrpar}, $T_\star$\,grid).  For our hydrodynamic models it
is moreover necessary to prescribe the stellar mass. We study two values,
$M_\star=67\,M_\odot$ and $55\,M_\odot$, corresponding to relatively large
Eddington factors $\Gamma_{\rm e}= 0.55$ and 0.67 (note that we are giving
here $\Gamma_{\rm e}\equiv\chi_{\rm e}L_\star/4\pi c G M_\star$ for a fully
ionized plasma).

The obtained mass loss rates are plotted in Fig.\,\ref{fig:mdot-wnl}.  We find
a strong dependence on $\Gamma_{\rm e}$ (or equivalently the $L/M$ ratio), {
  and} on the stellar temperature ($\dot{M} \propto T_\star^{-3.5}$).
{\changedA Note that both effects mark an important change in the properties of
  WR-type winds, compared to the radiatively driven winds of OB stars.  As we
  will discuss in more detail in Sect.\,\ref{sec:thick}, {{\changedB this is}} a
  direct consequence of the optically thick wind physics, where the location
  of the critical point is coupled to $\Gamma_{\rm e}$ and the local electron
  temperature $T_{\rm e}$.}

The synthetic spectra obtained from our calculations are consistent with
observed WNL spectra.  In Fig.\,\ref{fig:mdot-wnl} we have indicated four
specific objects which are in good qualitative agreement with our
models. They reflect the observed sequence of WNL spectral subtypes, starting
with WN\,6 at 55\,kK to WN\,9 at 31\,kK. Note that these objects show similar
line spectra as our models, but different luminosities and surface
compositions. Because of this, their actual mass loss rates may differ from
the values given in Fig.\,\ref{fig:mdot-wnl}. A detailed comparison with the
WN\,7 subtype WR\,22 is presented in the next section.

For a more comprehensive overview we compare our results to the stellar
parameters obtained by \citet{ham1:06} for the galactic WNL sample.  However,
because our computations are restricted to a fixed luminosity, we cannot
simply compare the obtained mass loss rates.  We thus employ the transformed
radius
\begin{equation}
\label{eq:rtrans}
R_{\rm t} = R_\star \left[\frac{\varv_\infty}{2500 \, {\rm km}\,{\rm s^{-1}}} 
\left/
\frac{\sqrt{D}\dot{M}}{10^{-4} \, {\rm M_\odot}\,{\rm yr^{-1}}}\right]^{2/3} 
\right.  ,
\end{equation}
which is a luminosity-independent measure of the wind density.  As outlined by
\citet{ham1:98}, models with the same $R_{\rm t}$ are scale invariant, i.e.,
for a given value of $T_\star$ and a given surface chemistry they show the
same line equivalent widths. The results are shown in Fig.\,\ref{fig:rtts}. In
the $R_t$--$T_\star$ plane our wind models reproduce the upper half of the
observed WNL sample, i.e., WNL stars with {low} wind densities.  {\changedA The
  large scattering in the observations makes it difficult to draw clear
  conclusions from this comparison. Note, however, that e.g.\ constant mass
  loss rates would show a clear trend with $R_t \propto R_\star \propto
  T_\star^{-2}$. The fact that the observations as well as our models show no
  clear trend thus supports our finding of increasing $\dot M$ with decreasing
  $T_\star$.}  The large part of the {high wind density} objects, which are
{not} reproduced by our models, belong to the enigmatic WN\,8 subtypes. These
stars show strong photometric variations. {\changedA Pulsations thus might play
  a role in their mass loss (see the discussion in Sect.\,\ref{sec:inst}).
  Moreover, most of them are located far away from stellar clusters, i.e.,
  their luminosities are {\changedD largely} unknown. They might thus belong to a
  different group of objects with much lower luminosities.}

{{\changedB In Fig.\,\ref{fig:vinf-wnl} we}} compare the obtained terminal
wind velocities with observations. {\changedA The galactic WR\,stars are known
  to show a trend of increasing $\varv_\infty$ for earlier spectral subtypes
  \citep{how1:92,nie1:02}. In Fig.\,\ref{fig:vinf-wnl} this trend is indicated
  by the grey shaded area, together with observed values for single WNL stars
  from \citet{ham1:06}.  Our models reproduce approximately the mean of the observed
  values.  However, although they show increasing $\varv_\infty$ with
  increasing $T_\star$, this trend is less pronounced than the observed one.
  Particularly around $T_\star = 50\,$kK our models tend to saturate whereas
  the observed values still increase. In Sect.\,\ref{sec:clump} we will show
  that this issue might be related to the detailed radial dependence of the
  clumping factor $D(r)$.}

\subsection{Test case: WR\,22}
\label{sec:wr22}

\begin{figure*}[]
\parbox[b]{0.99\textwidth}{\includegraphics[scale=0.93]{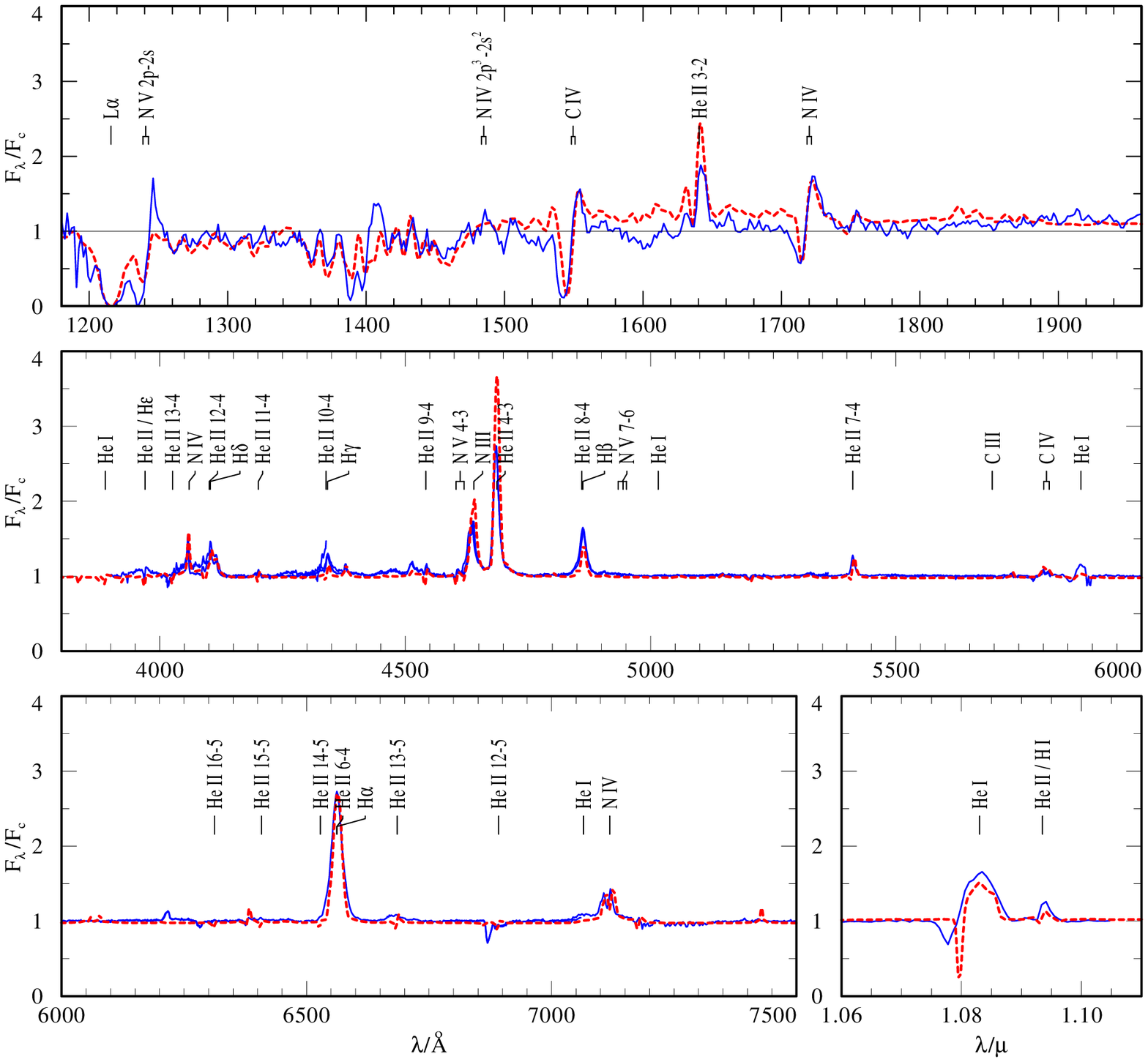}}  
\caption{
  WR\,22: Comparison of the synthetic spectrum from our hydrodynamic model
  (dashed line, red) with observations (blue). Fluxes are given in units of
  the model continuum flux. Note that {\em only} the optical and IR data are
  rectified by hand, whereas the observed UV fluxes are divided by the model
  continuum corrected for interstellar absorption and distance. The detailed
  model parameters are given in Table\,\ref{tab:wrpar}.}
  \label{fig:wr022}
\end{figure*}

\begin{figure}[]
 \parbox[b]{0.49\textwidth}{ 
 \includegraphics[scale=0.84]{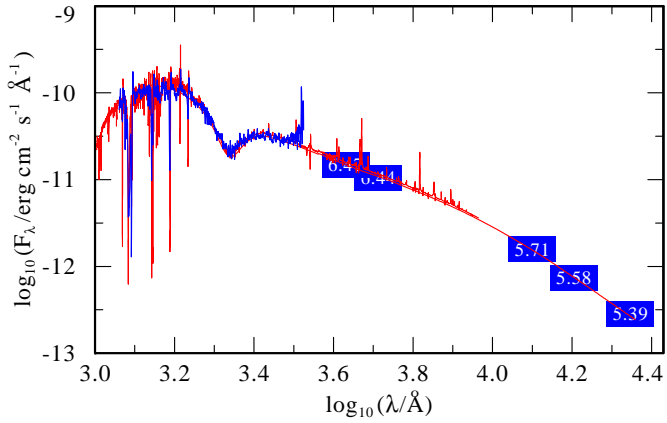}
 \caption{WR\,22: observed flux distribution in
      absolute units (blue), including optical + 2MASS IR photometry (blue
      boxes), compared with the model flux corrected for interstellar
      extinction {\changedD (red)}. The detailed model parameters are given in
      Table\,\ref{tab:wrpar}.  \label{fig:wr022_flux}}}
\end{figure}

To verify our assumption of high $L/M$ ratios for WNL stars we perform a more
detailed comparison with WR\,22, an eclipsing WR+O binary system {\changedC in
  Carina OB\,1,} which is classified as WN\,7h.  WR\,22 is an ideal target for
our purposes because mass estimates are available for the WN component.
{\changedA Moreover}, the O\,star component is so faint that at least 90\% of
the observed flux originates from the WR star, i.e., the WN spectrum is only
marginally contaminated.  From the radial velocity curve \citet{sch1:99}
derived a mass of $M_{\rm WR}\,\sin^3 i = 55.3\pm 7.3\,M_\odot$, whereas
\citet{rau1:96} obtained $71.7\pm 2.4\,M_\odot$, both with $M_{\rm WR}/M_{\rm
  O} = 2.7$.  The inconsistency is presumably due to the high eccentricity of
the system, and the {\changedD extremely high signal-to-noise ratio that is need
  to detect the O star absorption lines.}

In Fig.\,\ref{fig:wr022} we compare our best fitting synthetic spectrum to
observations of WR\,22. The {\changedC stellar} parameters are the same as for our
previous grid {\changedC models}, but with $T_\star = 44.7\,$kK and $X_{\rm
  H}=0.4$, as derived by \citet{ham1:06}. With our hydrodynamic models we
obtain a good spectral fit {\changedC for an Eddington parameter of $\Gamma_{\rm e}
  = 0.55$, corresponding to a mass of $78.1\,M_\odot$} (the detailed model
parameters are summarized in Table\,\ref{tab:wrpar}). {\changedC The resulting}
model reproduces the observed emission line spectrum reasonably well, although
the terminal wind velocity appears too small and the \HeI\ lines are slightly
too weak.  Nevertheless, the fit quality is more than satisfactory for our
present purposes. {\changedC The absolute flux distribution of WR\,22 (from IR to
  UV) is closely reproduced with a luminosity of $10^{6.3}\,L_\odot$ for} an
adopted distance modulus of 12.1\,mag {\changedC \citep{lun1:84}}, and an
interstellar extinction law according to \citet{fit1:99} with $E_{B-V}=0.42$
and $R=3.6$ (see Fig.\,\ref{fig:wr022_flux}).

{\changedC The mass of $78.1\,M_\odot$, as obtained from our models, is apparently
  in good agreement with the mass estimate by \citet{rau1:96}.  However, for a
  proper comparison we need to take into account the inclination of the
  system, its distance, and possiple uncertainties in our spectral analysis.}

{\changedC The inclination of the system is constrained by the detection of an
  eclipse near periastron, where the O star is obscured by the WR star
  \citep{gos1:91}. According to \citet{sch1:99} the periastron angle of the O
  star is 88\degr, and the orbit is highly eccentric with $e=0.598$.  No
  eclipse is detected near apastron, presumably because of the eccentricity of
  the orbit \citep{gos1:91}.  According to our models the radius of the
  stellar disk, i.e., the radius where $\tau_{\rm Ross}=2/3$, is only slightly
  larger than the stellar core radius: $R_{2/3}=1.056\,R_\star=25.0\,R_\odot$.
  Adopting $a \sin i = 2.31\,10^8\,{\rm km} = 13.3\,R_{2/3}$ from
  \citet{sch1:99}, we obtain a projected distance between both stars
  of $5.35\,R_{2/3}$ at periastron and $21.3\,R_{2/3}$ at apastron.  The
  eclipse detection at periastron thus constrains $\sin i$ to values above
  0.983 ($\sin^3 i \ge 0.950$). The non-detection at apastron limits $\sin i$
  to values below 0.9982 ($\sin^3 i \le 0.9946$), i.e., $\sin^3 i = 0.972\pm
  0.022$. Note that we have neglected the size of the O star in this estimate
  because a significant part of the O star is obscured but it is not clear
  whether the eclipse is full or partial.
  Applying our limits for $\sin^3 i$ we obtain $M_{\rm
    WR}=56.9\pm8.5\,M_\odot$ using the orbital data from \citet{sch1:99}, and
  $M_{\rm WR}=73.8\pm4.0\,M_\odot$ based on \citet{rau1:96}. Hence the
  inclination induces only small additional uncertainties in the mass
  estimate.  }

{\changedC The emission line strengths computed in our models depend strongly on
  the wind density, and thus on the obtained mass loss rates.  These} are
chiefly affected by $T_\star$ and $\Gamma_{\rm e}$.  While $T_\star$ can be reliably
determined from spectral line diagnostics, $\Gamma_{\rm e}$ depends on the
$L/M$ ratio and on the hydrogen abundance $X_{\rm H}$. {\changedC Again, $X_{\rm
    H}$ can be determined from spectral line ratios. Our models thus
  constrain the $L/M$ ratio, with an error of about 10\% (see
  Sect.\,\ref{sec:clump} for a discussion of additional systematic errors).
  The obtained value for $M_\star$ thus scales with $L_\star$, which depends
  on the adopted distance and interstellar extinction.  The extinction is
  determined within our spectral analysis from the continuum shape in the UV.
  Together with $T_\star$ this might introduce an additional error of about
  10\% in $L_\star$.  More critical is the adopted distance modulus for Carina
  OB1.  Values in the literature span a range between 11.8\,mag
  \citep{dav1:01} and 12.55\,mag \citep{mas1:93}, corresponding to an
  uncertainty of a factor of two in $L_\star$. The unknown distance of the
  system is thus the main source of uncertainty in our analysis.}

  We have seen that, with {\changedA an intermediate} distance modulus of
  12.1\,mag \citep[according to][]{lun1:84}, our model is in agreement with
  the mass determination by \citet{rau1:96}.  {\changedA For the low value of
    11.8\,mag, WR\,22 would become slightly fainter {\changedC
      ($10^{6.18}\,L_\odot$)} and our mass estimate would roughly scale down
    to $59\,M_\odot$, now in agreement with \citet{sch1:99}.  With 12.55\,mag
    we would obtain a very high luminosity of $10^{6.5}\,L_\odot$ which puts
    WR\,22 extremely close to the Eddington limit or even above {\changedC
      ($\Gamma_{\rm e} = 0.92$ for $74\,M_\odot$, and $\Gamma_{\rm e} = 1.24$
      for $55\,M_\odot$).  Note that in view of the additional b-b and b-f
      contributions to the mean opacity also $\Gamma_{\rm e} = 0.7$ would
      imply a super-Eddington luminosity (see Sect.\,\ref{sec:clump}). Such a
      large distance to WR\,22 thus seems to be very problematic.}}

At the present stage we conclude that WR\,22 {\changedA has a very high}
luminosity of the order of $10^{6.3}\,L_\odot$, and that it is located rather
close to the Eddington limit. {\changedA Chiefly because of the uncertain
  distance, the possible error in the luminosity is large. Nevertheless, mass
  estimates from radial velocity measurements support our assumption of large
  Eddington factors for WNL subtypes.  With the large distance determined by
  \citet{mas1:93}, WR\,22 would even exceed the Eddington limit.}

\subsection{Grid calculations for different metallicities}
\label{sec:zdep}

\begin{figure*}[t!]
  \parbox[b]{0.66\textwidth}{ \includegraphics[scale=0.46]{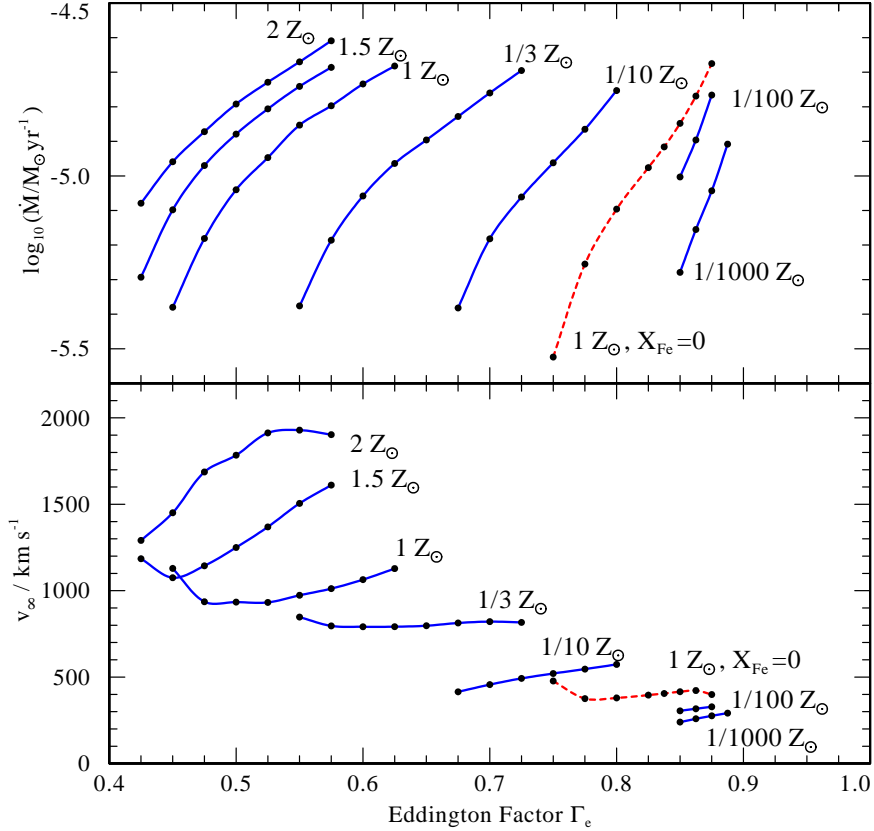} }
\parbox[b]{0.33\textwidth}{
\caption{ WNL star mass loss in the range from
        $Z_\odot/1000$ to $2\,Z_\odot$.: Mass loss rates (top) and terminal
        wind velocities (bottom), as obtained from our hydrodynamic grid
        models, are plotted vs.\ the Eddington factor $\Gamma_{\rm e}$ (the
        models are computed for a fixed value of $L_\star =
        10^{6.3}\,L_\odot$, the variation of $\Gamma_{\rm e}$ thus effectively
        corresponds to a variation of the stellar mass). Model parameters are
        given in Table\,\ref{tab:wrpar}. The solid blue lines indicate model
        series where $\Gamma_{\rm e}$ is varied for a given value of $Z$.
        Note that strong WR-type mass loss occurs over the whole range of $Z$,
        if $\Gamma_{\rm e}$ approaches unity. Moreover, the models show a
        clear trend of decreasing $\varv_\infty$ with decreasing $Z$. The
        dashed red lines indicate models with zero Fe, but solar-like CNO
        abundance. They show that strong WR-type mass loss is also possible
        for extremely metal-poor stars due to the self-enrichment with primary
        nitrogen.  \label{fig:wnl-z} }}
\end{figure*}

{\changedA The wind acceleration in our models predominantly originates from Fe
  line opacities. We thus expect a strong dependence of the mass loss on
  metallicity.  To investigate this dependence we have prepared a grid of
  hydrodynamic WNL models for a wide range of $Z$.  Because we already know
  about the fundamental importance of the Eddington limit for WR-type winds we
  have introduced $\Gamma_{\rm e}$ as an additional free parameter.} The stellar
core parameters ($L_\star$, $T_\star$, $R_\star$), and the surface abundances
$X_{\rm H}$ and $X_{\rm He}$ are {\changedA adopted from our WR\,22 model in
  the previous section}.  $M_\star$ is adjusted to match the desired values of
$\Gamma_{\rm e}$, and $Z$ is varied by scaling the {\changedA surface mass fractions}
$X_{\rm C}$, $X_{\rm N}$, $X_{\rm O}$, $X_{\rm Si}$, and $X_{\rm Fe}$ (see
Table\,\ref{tab:wrpar}, Grid\,2).

The results are presented in Fig.\,\ref{fig:wnl-z}. {\changedA The solid (blue)
  curves indicate model series for different {\em initial} metallicities,
  i.e., with CNO abundances corresponding to a secondary helium and nitrogen
  enrichment by the CNO-process.} As expected, the models show a strong
{\changedA $Z$-dependence. However, our second free parameter $\Gamma_{\rm e}$}
turns out to be at least equally important.  In fact it seems that the
proximity to the Eddington limit is the {primary reason} for the enhanced mass
loss of WR stars. {\changedA Remarkably, the} high WR-type mass loss rates can
{\changedA even be maintained} for very low $Z$, if the stars {\changedA are}
close enough to the Eddington limit.

The dashed (red) curve in Fig.\,\ref{fig:wnl-z} indicates models with $X_{\rm
  Fe}=0$ and $X_{\rm Si} =0$ but $X_{\rm CNO}=Z_\odot$ (see
Table\,\ref{tab:wrpar}, Grid\,3).  These models {{\changedB show}} to what extent
the {\em primary} production of nitrogen may affect the mass loss at different
metallicities.  Such an enhancement is expected for fast rotating stars at
extremely low metallicities \citep[see][]{mey1:06}.  Interestingly, these
models show a similar behavior to the Fe-rich models.  The solar-like CNO mass
fraction produces a stellar wind {\changedA that is similar} to an Fe-rich
model with $1/50\,Z_\odot$.  This means that the potential mixing of primary
nitrogen to the surfaces of {\changedA metal-poor stars} considerably affects
the expected mass loss {\changedA rates}.

In the bottom panel of Fig.\,\ref{fig:wnl-z} we show the terminal wind
velocities obtained from our models.  First, {\changedA the} models clearly
predict decreasing wind velocities with decreasing $Z$.  This finding is
{\changedA in agreement with} observations although the observational evidence
is not as clear as our prediction \citep[see e.g.][]{con1:89,cro2:00}.
{\changedA Second, the terminal velocities tend to stay constant or even
  increase with $\dot{M}$.  This behaviour again indicates important
  qualitative differences to OB star winds. For OB stars the well-established
  wind momentum-luminosity relation predicts the opposite behaviour,
  i.e., decreasing wind velocities for increasing mass loss
  \citep[][]{kud1:99}. A closer inspection of our models shows that the
  changes in the velocity structure are related to changes in the wind
  ionization. For cases where an increase of $\dot M$ leads to recombination
  of Fe, the velocity increases in the recombination region due to the newly
  exposed line opacities.  Models with a similar ionization structure, on the
  other hand, show similar wind velocities.}  Note that our assumption of a
constant Doppler broadening ($\varv_{\rm D} = 100\kms$, see
Sect.\,\ref{sec:mpar}) might affect the results for extremely low $Z$, because
for these cases the wind velocities are of the same order of magnitude as
$\varv_{\rm D}$.

{\changedA
\subsection{A mass loss recipe for WNL stars}
\label{sec:recipe}

\begin{figure}[t!]
\vskip0.425cm
\noindent
\parbox[b]{0.49\textwidth}{\includegraphics[scale=0.4]{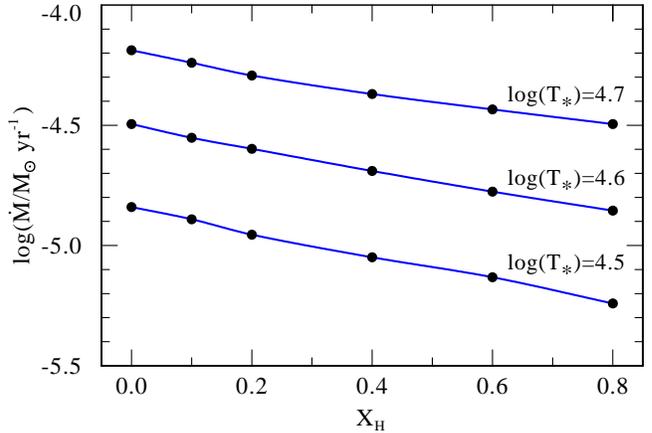}}
\caption{\changedA The dependence of WNL star mass loss on the hydrogen
  abundance $X_{\rm H}$: for otherwise fixed stellar parameters the mass loss
  depends only weakly on $X_{\rm H}$, with $\dot M \propto {X_{\rm
      H}}^{0.45}$. Model computations for three different values of $T_\star$
  and fixed values of $\Gamma_{\rm e}=0.55$ and $L_\star=10^{6.3}L_\odot$ are
  shown.  Note that the stellar mass is varied to keep $\Gamma_{\rm e}$
  constant.}
\label{fig:XH}
\end{figure}

\begin{figure}[t!]
\vskip0.425cm
\noindent
\parbox[b]{0.49\textwidth}{\includegraphics[scale=0.4]{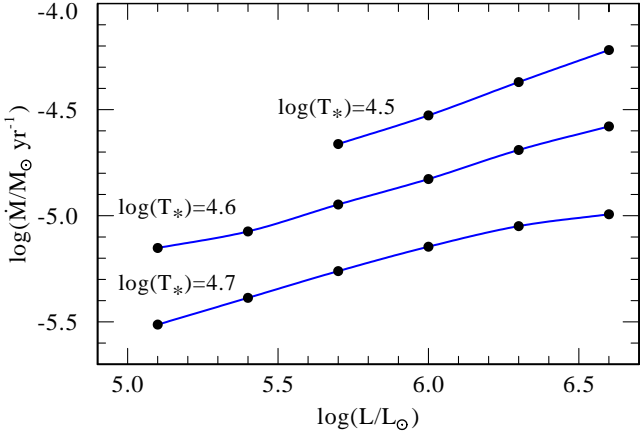}}
\caption{\changedA The dependence of WNL star mass loss on luminosity: for
  otherwise fixed stellar parameters the mass loss depends relatively weakly
  on $L_\star$, with $\dot M \propto L_\star^{0.42}$. Model computations for
  three different values of $T_\star$ are shown. Note that the Eddington
  factor is fixed to $\Gamma_{\rm e}=0.55$ by keeping the $L/M$ ration
  constant.  }
\label{fig:lumi}
\end{figure}

A primary goal of our present work is to provide a recipe that describes the
mass loss of WNL stars, dependent on the fundamental stellar parameters
$M_\star$, $T_\star$, $L_\star$ and the surface mass fractions $X_{\rm H}$,
$X_{\rm He}$, and $Z$. In the preceding sections we emphasized
that, for fixed $L_\star$, the mass loss is chiefly determined by the
proximity to the Eddington limit, i.e., by the Eddington factor $\Gamma_{\rm
  e}$. We thus expect that the main influence of the surface mass fractions
$X_{\rm H}$ and $X_{\rm He}$ is due to the number of free electrons, which
enters the Eddington parameter for a fully ionized plasma in the form
$\Gamma_{\rm e} = 10^{-4.813} (1+X_{\rm H})\, L_\star/M_\star$. In
Fig.\,\ref{fig:XH} we show that for fixed $\Gamma_{\rm e}$ the influence of
the light elements is indeed only moderate, with $\log(\dot M)\propto
-0.45\,X_{\rm H}$. As we will discuss in Sect.\,\ref{sec:comp}, the slight
decrease of $\dot M$ with increasing hydrogen abundance is due to the
influence of the \ion{He}{ii} continuum in the critical layers.

We present in Fig.\,\ref{fig:lumi} a series of
models with varied $L_\star$.  Again, the stellar parameters are adopted from
WR\,22. $\Gamma_{\rm e}$ is fixed to a value of 0.55, and $L_\star$ and
$T_\star$ are varied over the relevant parameter range (see
Table\,\ref{tab:wrpar}). Notably, the resultant mass loss rates depend only
weakly on $L_\star$, with $\dot M \propto L_\star^{0.42}$. Note that this
again denotes an important difference to OB star winds for which a much
stronger dependence is expected (see the discussion in
Sect.\,\ref{sec:thick}).

We combine our results in the following form.  For the dependence on
$T_\star$ we use the scaling relation from Sect.\,\ref{sec:test} with $\dot M
\propto T_\star^{-3.5}$, and for the dependence on ${X_{\rm H}}$ and $L_\star$
we use $\dot M \propto {X_{\rm H}}^{0.45} L_\star^{0.42}$ from this section.
For the coupled dependence on $\Gamma_{\rm e}$ and $Z$ from
Sect.\,\ref{sec:zdep} we use a more complex description of the form $\dot M
\propto (\Gamma_{\rm e} - \Gamma_0)^\beta$, where the $Z$-dependent parameters
$\Gamma_0(Z)$ and $\beta(Z)$ are determined by a multi-dimensional fit
procedure.  The resultant mass loss recipe is in good agreement with our grid
computations in Fig.\,\ref{fig:wnl-z}, with an accuracy better than 0.05\,dex
except for two data points with a maximum deviation of 0.1\,dex.  Combining
all these relations, we arrive at a mass loss recipe of the form
\begin{equation}
  \label{eq:massloss1}
  \begin{array}{l}
    \log\left(\dot{M}_{\rm WNL}/\msunpyr\right) = -3.763\\
    \;\;\;\;\;\;
    +\, \beta \cdot \log\left(\Gamma_{\rm e} - \Gamma_0\right)
        - 3.5\cdot\left(\log (T_\star/{\rm K}) - 4.65\right)\\
    \;\;\;\;\;\;
    +\, 0.42\cdot\left(\log (L_\star/L_\odot)-6.3\right)
      - 0.45\cdot\left(X_{\rm H} - 0.4\right),\\
  \end{array}
\end{equation} 
or
\begin{equation}
  \label{eq:massloss}
  \begin{array}{l}
    \log\left(\dot{M}_{\rm WNL}/\msunpyr\right) = 10.046 \\
    \;\;\;\;\;\;
    +\, \beta \cdot \log\left(\Gamma_{\rm e} - \Gamma_0\right)
        - 3.5 \log (T_\star/{\rm K})\\
    \;\;\;\;\;\;
    +\, 0.42 \log (L_\star/L_\odot) - 0.45\, X_{\rm H},
  \end{array}
\end{equation} 
with
\begin{eqnarray}
  \label{eq:fit2}
  \beta(Z) &=& 1.727 + 0.250 \cdot \log(Z/Z_\odot), \\
  \label{eq:fit3}
  \Gamma_0(Z) &=& 0.326 - 0.301 \cdot \log(Z/Z_\odot) - 0.045 \cdot \log(Z/Z_\odot)^2.
\end{eqnarray}
This recipe is only valid for optically thick winds that are driven by the
cool Fe opacity-peak. It is thus only applicable in a limited range of stellar
temperatures. $T_\star$ should never exceed a value of $\approx$\,70\,kK
because then the opacity minimum between the hot and the cool Fe-peaks is
reached. On the cool side, for $T_\star < 30$\,kK, our models reach a limit
where the radiative force exceeds gravity in the deep hydrostatic layers due
to the hot Fe-peak.  At this point the region of LBV-type instabilities might
be reached (see the discussion in Sect.\,\ref{sec:inst}). 

{{\changedB According to Eqs.\ (\ref{eq:massloss1}) and (\ref{eq:massloss}) the WNL
    mass loss rates would go to zero when $\Gamma_{\rm e}$ approaches
    $\Gamma_0$.  Our recipe of course loses validity before this
    point, when the winds become optically thin.  For our present grid models
    this happens at a mass loss rate of $\log(\dot M/\msunpyr) \approx -5.5$
    when $\Gamma_{\rm e}-\Gamma_0\approx 0.1$. Note that this criterion should
    roughly depend on the wind density at the sonic point. We thus expect the
    limiting mass loss rate, above which our models are valid, to be of the
    order of $\log(\dot M_{\rm lim}/\msunpyr) \approx -5.5 + 2
    \log(R_\star/23.7\,R_\odot)$. Note that Eqs.\ (\ref{eq:fit2}) and
    (\ref{eq:fit3}) which describe our parameters $\Gamma_0(z)$ and
    $\beta(z)$}} are strictly limited to the investigated parameter range of
$2\,Z_\odot < Z < 10^{-3}Z_\odot$.  }

\section{The properties of WR-type stellar winds}
\label{sec:wdrv}

{\changedA In the present section we discuss the properties of WR-type stellar
  winds as opposed to the thin winds of OB stars (Sect.\,\ref{sec:thick}).  In
  this context we discuss possible error sources, with emphasis on the unknown
  turbulent velocities and wind clumping factors (Sect.\,\ref{sec:clump}).
  {{\changedB Moreover,}} we compare our results with other mass loss predictions
  for WR and OB stars (Sect.\,\ref{sec:comp}).

\subsection{Optically thick stellar winds}
\label{sec:thick}

In Sects.\,\ref{sec:test} and \ref{sec:recipe} we have seen that our WR wind
models are qualitatively different from OB-type stellar winds.  In particular
we find a strong dependence on the Eddington factor $\Gamma_{\rm e}$, a steep
increase of $\dot M$ with decreasing effective temperature ($\dot M \propto
T_\star^{-3.5}$), and a weak dependence on the stellar luminosity ($\dot M
\propto L_\star^{0.4}$). Our results from Sect.\,\ref{sec:zdep} even suggest a
violation of the wind momentum-luminosity relation, which is extremely
well-established for OB star winds {\changedC \citep[see,
e.g.,][]{kud1:99,mok1:05,mok2:07}}.  Instead, the velocity structure in our
models seems to be dominated by ionization effects.

Within the classical CAK theory for OB star winds \citep{cas1:75} a mass loss
relation of the form $\dot M \propto M_{\rm eff}^{(\alpha-1)/\alpha}
L_\star^{1/\alpha}$ is expected. As explained in Sect.\,\ref{sec:numerics},
the parameter $\alpha$ characterizes the response of the radiative
acceleration to the velocity gradient, and {\changedC typically} lies around
$\alpha$\,$\approx$\,$0.6$. $M_{\rm eff}$ denotes the effective stellar mass,
reduced by the contribution of the radiative acceleration on free electrons
$M_{\rm eff}=M_\star (1-\Gamma_{\rm e})$. In contrast to our WR models, the
expected dependence on the stellar mass is thus rather weak, while the
luminosity dependence is comparatively strong. Moreover, OB star winds do not
have a pronounced wind ionization structure, apart from a sudden change in
ionization around spectral type B1 \citep[the bistability jump,
see][]{lam1:95}. The ionization thus only plays a minor role in the wind
dynamics and is usually expressed by a small correction to the $\alpha$
parameter \citep[see][]{abb1:82}.

As outlined in Sect.\,\ref{sec:numerics}, the $\alpha$ parameter is determined
numerically within our models and enters the hydrodynamic solution scheme as a
depth-dependent quantity. Compared with the typical values for OB stars
($\alpha$\,$\approx$\,$0.6$) our present models show remarkably low values
{{\changedB ($\alpha$\,$\approx$\,$0.2$)}}.  This result is in line with our
previous finding of $\alpha$\,$\approx$\,$0$ for the extremely thick winds of
WC stars \citep[][see also the discussion therein]{gra1:05}. It seems that
$\alpha$\,$\rightarrow$\,$0$ for thick winds. Such a behaviour is expected for
the deep atmospheric layers, where the diffusion approximation is applicable.
For the outer wind such low values again indicate important qualitative
differences to 'classical' line-driven winds.  At the present stage we
attribute this effect to two mechanisms, which are expected for the case of
WR-type winds, and which are both detectable in our models: the dominance of
weak spectral lines, and massive line overlap.

First, the high density of WR winds leads to an over-population of excited
energy levels by recombination cascades. This results in an increase of the
number of optically thin lines.  Note that $\alpha$ also {{\changedB
    reflects}} the relative contribution of optically thick spectral lines to
the overall wind driving \citep[see, e.g.][]{pul1:00}. A dominance of
optically thin lines thus leads to a decrease of $\alpha$, or in other words
to the formation of a pseudo-continuum of thin lines. Second, the resulting
ionization structure leads to non-local line overlaps. {\changedC In such a
  situation it can happen that an additional Doppler-shift causes a reduction
  of the radiative force because of the increased line-shadowing.}
Both effects result in a dominance of ionization effects over the usual CAK
wind physics, as is observed in our models.

Alternative theories for optically thick stellar winds have been proposed by
\citet{pis1:95} and \citet{nug1:02}.  Both works assume that the conditions at
the critical point depend on the Rosseland mean opacity, and they show that
the observed mass loss rates of WR stars are in agreement with this
assumption. In the following we will show that the scaling relations obtained
from our present models can be understood within this {{\changedB framework}}. They
reflect the dominant role of the wind optical depth for WR-type mass loss,
which we identify as the basic reason for the qualitative differences to OB
star winds.  Note, however, that the conditions within our models do not
strictly reflect the limit of optically thick winds.  Our present WNL models
show only moderate optical depths, so that their mass loss actually reflects a
{{\changedB case}} in between the optically thick and the optically thin limit.

The basic assumption of the optically thick wind {{\changedB approach}} is that the
radiation field at the critical point of the wind flow can be described in the
diffusion limit {{\changedB \citep[see][]{nug1:02}}}. It follows that 1) the sonic
point becomes the critical point of the equation of motion $\varv_{\rm
  c}=a(T_{\rm c})$, 2) a specific (critical) value of the Rosseland mean
opacity must be reached at the sonic point $\chi_{\rm c} = 4\pi c G M_\star /
L_\star$, and 3) the mass loss rate is connected to the density and the
temperature at this point via the equation of continuity $\dot M = 4\pi
R_\star^2 \rho_{\rm c} \varv_{\rm c}$.

The importance of the $L/M$ ratio for this kind of mass loss follows directly
from condition 2), which means that the radiative acceleration just
balances the gravitational attraction at the critical point.
For the formation of an optically thick wind it is thus {{\changedB
    mandatory}} that the $L/M$ ratio and $\chi_{\rm c}$ are large enough,
i.e., the star must be close to the Eddington limit and the mean opacity at
the critical point must be large. Furthermore, the mean opacity must increase
outward to obtain an outward-directed net acceleration above the critical
point. These conditions restrict the temperature $T_{\rm c}$ at the critical
point to limited temperature ranges connected to the occurrence of
the two Fe opacity-peaks. With $T_{\rm c} = 30$--45\,kK our present WNL models
are driven by the cool Fe-peak opacities.  Our previous WCE model with $T_{\rm
  c} = 199$\,kK, on the other hand, clearly belonged to the regime {\changedC of
  the hot Fe-peak} \citep[see][]{gra1:05}.

The necessity to reach certain values of $T_{\rm c}$ marks a basic change of
paradigm with respect to OB star winds.  {\changedC The temperature at large
  optical depth scales as $T^4 \propto T_\star^4 \tau$.  Hence} large wind
optical depths $\tau_{\rm c}$ are needed to reach large values of $T_{\rm c}$.
For our case, where a specific temperature $T_{\rm c}$ must be reached at
$\tau_{\rm c}$ this means that $\tau_{\rm c} \propto T_\star^{-4}$.

We found a very strong dependence on $T_\star$ for our models with
fixed $\Gamma_{\rm e}$ in Sect.\,\ref{sec:test}.  This behaviour is related to
the scaling properties of $\tau_{\rm c}$.  If we assume that $\tau_{\rm c}$
scales with $\rho_{\rm c}$ and the density scale height $H$, we get $\tau_{\rm
  c} \propto \rho_{\rm c} H \propto \rho_{\rm c} R_\star^2 / M_{\rm eff}
\propto \rho_{\rm c} R_\star^2 / M$. From this it follows directly that
$\tau_{\rm c} \propto \dot M / M$ because $\dot M = 4\pi R_\star^2 \rho_{\rm
  c} \varv_{\rm c} \propto R_\star^2 \rho_{\rm c}$, for {{\changedB the}} given
value of $\varv_{\rm c}=a(T_{\rm c})$.  Finally we get $T_\star^{-4} \propto
\tau_{\rm c} \propto \dot M / M$, or $\dot M \propto T_\star^{-4} L$ for fixed
$L/M$.
 
This scaling relation indeed predicts a strong dependence on $T_\star$ and a
relatively weak dependence on $L$ for optically thick winds.  It {{\changedB is
    caused}} by the scaling properties of $\tau_{\rm c}$ with the density
scale height $H$, and the requirement that a certain temperature regime is
reached at the critical point.  The formation of optically thick winds {{\changedB
    is thus strongly related to the extension of the deep atmospheric layers
    (i.e., the increase of $H$)}} close to the Eddington limit.

With $\dot M$\,$\propto$\,$T_\star^{-3.5} L^{0.4}$ our computational models
show a similar dependence on $T_\star$ and an even weaker dependence on $L$.
The differences are presumably due to the idealized assumptions in our
analytic derivation.  Nevertheless, we conclude that our models show a
qualitative behaviour that is very different from OB star winds, but can be
understood {{\changedB in the framework of the optically thick wind approach}}.
Note that the specific properties of optically thick stellar winds originate
from the deep atmospheric layers, and that an exact modelling of these layers
is required to obtain qualitatively correct results.

\subsection{The influence of wind clumping and micro-turbulence}
\label{sec:clump}
\begin{figure}[t!]
\parbox[b]{0.49\textwidth}{\includegraphics[scale=0.41]{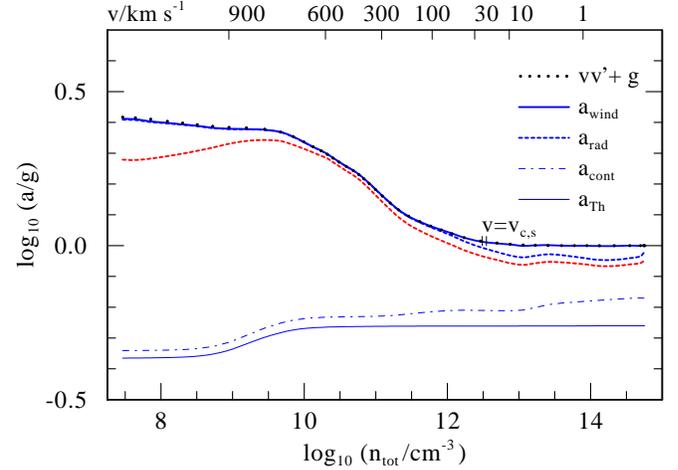}}
\caption{\changedA WR\,22: wind acceleration in units of the local gravity $g$.  The
  total atomic density is given as depth index. The blue curves indicate the
  acceleration within our WR\,22 model from Sect.\,\ref{sec:wr22}.  Indicated
  are the radiative acceleration on free electrons ($a_{\rm Th}$), on continua
  ($a_{\rm cont}$), the total radiative acceleration including lines ($a_{\rm
    rad}$), and the total wind acceleration including the pressure term
  ($a_{\rm wind}$). The red line indicates the radiative acceleration within our
  test model with a reduced Doppler broadening velocity $\varv_{\rm D}$\,$=$\,50\,km/s.
\label{fig:acc}
}
\end{figure}

Apart from the work by \citet{sch1:97}, the influence of wind clumping has
been largely ignored in previous studies of radiatively driven winds.  In our
own work we have demonstrated that clumping increases the radiative
acceleration in O supergiant \citep{gra1:02,gra2:03} and WR star winds
\citep{gra1:05}.
The specific choice of the clumping factor $D(r)$ thus has an influence on our
results. In Sect.\,\ref{sec:mpar} we have described how $D(r)$ is parametrized
within our models. This parametrization is based on observational grounds.
Clumping diagnostics {\changedC are} however very rough and restricted to the
formation region of the diagnostic features.  In principle, $D(r)$ thus could
be varied in a certain range. In the present section we discuss to what
extent this would affect our results. In addition, we investigate the
influence of the Doppler broadening velocity $\varv_{\rm D}$.
In our models $\varv_{\rm D}$ is usually set to a fixed, highly supersonic
value of $\varv_{\rm D}$\,$=$\,100\,km/s. This choice is motivated by the
observed broadening velocities in P-Cygni type line profiles
\citep[e.g.][]{ham1:81}, and by the fact that our CMF radiative transfer is
limited to relatively large values of $\varv_{\rm D}$ (see
Sect.\,\ref{sec:mpar}).  Again, we have a certain freedom in the choice of
$\varv_{\rm D}$ and we want to quantify its influence on our results.

We start with Fig.\,\ref{fig:acc} where we show the wind acceleration within
our model for WR\,22.  In this logarithmic diagram we plot the wind
acceleration $a_{\rm wind}$ in units of the local gravity $g$, vs.\ the total
atomic number density $n_{\rm tot}$ as depth index.  The plot covers the whole
model atmosphere from the hydrostatic layers where $\log(a_{\rm wind}/g)=0$,
through the critical point, to the outer boundary at $r=1000\,R_\odot$.  The
critical point (with $\varv_{\rm c}=23.3\,{\rm km/s}$) and the sonic point
(with $\varv_{\rm s}=20.8\,{\rm km/s}$) are both located very close together
at the point where $a_{\rm rad} \approx g$, in agreement with our condition 2)
from the previous section.  In the hydrostatic layers below the sonic point
(at densities {\changedC above} $\log(n_{\rm tot}/{\rm cm}^{-3})\approx 12.5$), the
radiative acceleration $a_{\rm rad}$ (indicated by the dashed blue line) lies
only slightly below $g$, with $\log(a_{\rm rad}/g) \approx -0.05$. This shows
that, although with $\Gamma_{\rm e}=0.55$ we are still away from the Eddington
limit, our model atmosphere is in fact rather close to instability (with a
'true' Eddington parameter, including all opacity sources, of $\Gamma \approx
0.9$).

{{\changedB Being}} a late spectral subtype, WR\,22 is in a temperature regime
where the cool Fe-peak opacities dominate. In Fig.\,\ref{fig:acc}, however, we
can see that close to the critical point $\approx$\,70\,\% of the radiative
force is due to continua.  Nevertheless, the Fe line opacities play the
dominant role in determining the mass loss because they are responsible for
the {\em increase} of $a_{\rm rad}$ which finally leads to the crossing of the
Eddington limit at the critical point. The fact that the line contribution in
the deep layers is so small compared to continua (which are dominated by
Thomson scattering) indicates that mostly weak lines are responsible for the
radiative driving, i.e., the sub-critical layers are supported by a
pseudo-continuum of millions of weak Fe-group transitions.

From Fig.\,\ref{fig:acc} it also becomes clear why our models are extremely
sensitive to changes of $a_{\rm rad}$ close to the critical point. The dashed
red line indicates the radiative acceleration within a model with exactly the
same atmospheric structure but with a reduced broadening velocity of
$\varv_{\rm D}$\,$=$\,50\,km/s.  This causes a rather small reduction of
$a_{\rm rad}$ (by roughly $-0.025$\,dex) in the deep atmospheric layers. In the
outer wind region the relative effect is somewhat larger ($-0.1$\,dex), the
absolute wind velocity is however hardly affected because the wind is already
close to its terminal speed.  Although the effects in the deep layers are
small it is obvious that even this small reduction of $a_{\rm rad}$ might
cause a significant shift of the critical point towards smaller densities
(i.e., smaller $\dot{M}$).  A corresponding adjustment of $M_\star$, however,
brings the star back to the same distance from the Eddington limit, and should
thus restore the previous value of $\dot{M}$.

\begin{table}[]
\changedA 
  \begin{center}
  \begin{tabular}{llllll} 
    \hline \hline 
    \rule{0cm}{2.2ex}Model & WR\,22 & $\varv_{\rm D1}$ & $\varv_{\rm D2}$
    & $D$\,1 & $D$\,2\\
    \hline
    \rule{0cm}{2.2ex}$M_\star / M_\odot$  
                           & 78.1   & 78.1   & 73.5   & 78.1 & 78.1 \\
    $\Gamma_{\rm e}$       & 0.55   & 0.55   & 0.585  & 0.55 & 0.55 \\
    $\varv_{\rm D} / \kms$ & 100    & 50     & 50     & 100  & 100  \\
    $D_{\rm max}$          & 10     & 10     & 10     & 40   & 40   \\
    $\tau_1$               & 0.7    & 0.7    & 0.7    & 0.7  & 0.7 \\
    $\tau_2$               & 0.35   & 0.35   & 0.35   & 0.35 & 0.175 \\
    \hline 
    \rule{0cm}{2.2ex}$\dot{M} / \msunpyr$ & $10^{-4.85}$ & $10^{-4.99}$
                                          & $10^{-4.87}$ & $10^{-4.61}$ & $10^{-4.80}$ \\
    $\vinf / \kms$                        & 974  & 932  & 938  & 1859 & 1764 \\ 
    $R_\mathrm{t} / R_\odot$              & 21.7 & 25.9 & 21.7 & 14.5 & 18.6 \\ 
    $\eta$                                & 0.34 & 0.23 & 0.31 & 1.12 & 0.69 \\ 
    \hline 
  \end{tabular}
  \end{center}
  \caption{Model parameters and resulting wind parameters for our test models
    from Sect.\,\ref{sec:clump}, compared to the WR\,22 model from
    Sect.\,\ref{sec:wr22}. For models $\varv_{\rm D1}$ and $\varv_{\rm D2}$
    the Doppler broadening velocity has been reduced to $\varv_{\rm D}$\,$=$\,50\,km/s.
    For models $D1$ and $D2$ the wind clumping has been increased by a factor
    of four. 
  }
  \label{tab:vdtest} 
\end{table}

Test computations indeed confirm this behaviour.  In Table\,\ref{tab:vdtest}
we compare our WR\,22 model with full hydrodynamic models for which
$\varv_{\rm D}$ has been reduced. For model $\varv_{\rm D1}$ all parameters
apart from $\varv_{\rm D}$ are kept fixed, and $\dot{M}$ is indeed reduced by
0.24\,dex.  For model $\varv_{\rm D2}$ we have compensated this effect by a
reduction of the stellar mass (0.026\,dex) which indeed brings the star back
to the previous state.  Our results do
depend on the assumed microphysics, because the location of the critical point
strongly depends on the ratio $a_{\rm rad}/g$ in the deep atmospheric layers.
For the same reason our predicted mass loss rates depend strongly on the
Eddington factor $\Gamma_{\rm e}$. The relatively large value of $\varv_{\rm
  D}$\,$=$\,100\,km/s that we have chosen for our computations might thus have
a significant effect on $\dot M$. However, it only causes a systematic shift
of our results towards lower stellar masses, by $\sim 5\,\%$.  Because the
actual conditions in the sub-photospheric layers of WR\,stars are unknown, it
is difficult to say if our assumption is correct.

The influence of wind clumping turns out to be even more important. In the
last two columns of Table\,\ref{tab:vdtest} we summarize the results of two
test computations where the maximum clumping factor $D_{\rm max}$ has been
scaled up by a factor of 4. Note that the spectroscopic properties of
WR\,stars are determined by the product $\dot M\sqrt{D}$ \citep[for details,
see][]{ham1:98}. An increase of the clumping factor by a factor of four thus
approximately mimics {the opacity} of a wind with two times higher density.
{\changedD Consequently,} $a_{\rm rad}$ also should scale with $\sqrt{D}$. Note that such a
scaling relation is only expected to hold strictly for {{\changedB the}}
populations of excited {{\changedB energy}} levels which are dominated by
recombination processes.

Our test computations (models $D1$ and $D2$ in Table\,\ref{tab:vdtest}) show
that the expected scaling of $a_{\rm rad} \propto \sqrt{D}$ is fulfilled with
an astonishing precision. Both test models with $D_{\rm max}$\,$=$\,40 show
nearly exactly twice the terminal wind speed as our $D_{\rm max}$\,$=$\,10
models. Again, this is a hint that weak line transitions from excited
(recombination-dominated) levels play a major role in the wind driving of WR
stars. Moreover, the radial dependence of the wind clumping is also important.
With our standard prescription of a $\tau$-dependent {{\changedB
    clumping factor $D$, which rises from $D$\,$=$\,$1$ to $D_{\rm max}$
    between Rosseland optical depth $\tau_1=0.7$ and $\tau_2=0.35$ (see
    Sect.\,\ref{sec:mpar}),}} the increase of $D_{\rm max}$ also leads to a
significant increase of $\dot M$ (model $D1$ in Table\,\ref{tab:vdtest}).  The
reason for this is that in our prescription of $D(\tau)$ the wind clumping
already sets in below the critical point, and thus affects the critical
condition.  For model $D2$ we have set $\tau_2$ to a value of 0.175. In this
way we have modified the growth rate of $D(r)$ {{\changedB such}} that the
conditions at the critical point are similar to our WR\,22 model.  Again, the
mass loss reacts very sensitively to the change of $D(r)$ and returns to the
{{\changedB same value as the WR\,22 model.}}

We conclude that different assumptions in the microphysics of our models may
indeed cause significant changes in the predicted mass loss {\changedD rates}, if the
region around the critical point is affected. However, these effects can
always be compensated by a slight re-adjustment of the prescribed stellar
mass. For the wind clumping we find that $a_{\rm rad}$ scales with $\sqrt{D}$.
An increase of $D$ in the outer wind layers only leads to an increase of the
terminal wind speed by a factor $\sqrt{D}$.  The possible discrepancies of our
predicted wind speeds with observations (Sect.\,\ref{sec:zdep}) thus might be
resolved by a strong increase of $D(r)$ in the outer wind.

\subsection{Comparison with other mass loss predictions}
\label{sec:comp}

\begin{figure}[t!]
  \parbox[b]{0.49\textwidth}{\includegraphics[scale=0.41]{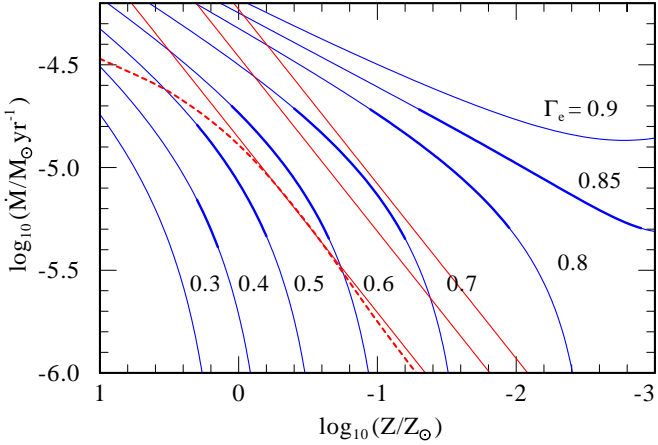}}
  \caption{\changedB The $Z$-dependence of WNL star mass loss: comparison of
    different mass loss predictions. The solid blue lines indicate the
    $Z$-dependent mass loss rates according to
    Eqs.\,(\ref{eq:massloss})--(\ref{eq:fit3}) for fixed values of
    $\Gamma_{\rm e}$\,$=$\,0.3--0.9 (note that only the thick parts of the
    blue curves are in the region where grid models have actually been
    computed). They are compared to the mass loss predictions by
    \citet[][solid red lines]{vin1:01} for $\Gamma_{\rm e}$\,$=$\,0.3, 0.6,
    and 0.9, and the mass loss relation for WNL stars by \citet[][dashed red
    line]{vin1:05}, which was obtained with $\Gamma_{\rm e} = 0.34$.
  \label{fig:zdep}
}
\end{figure}

In the last few years an impressive agreement between theoretical mass loss
predictions for OB stars and observations has been achieved (but see the
discussion in Sect.\,\ref{sec:dcl}).  On the theoretical side, models by
\citet{vin1:00,vin1:01}, \citet{pau1:01}, and \citet{krt1:04} agree well.
Observationally, these results are confirmed on a large scale by the
VLT-FLAMES survey of massive stars \citep{mok1:05,mok1:07}. For the
$Z$-dependence an exponential relation of the form $\dot M \propto Z^{\gamma}$
is generally accepted.  {\changedC \citet{vin1:01} find an exponent of $\gamma
  \approx 0.85$ for a broad range of OB star parameters. Note that this
  exponent is only valid if $\varv_\infty$ is kept fixed. It thus needs to be
  corrected for the dependence of $\varv_\infty(Z)$, resulting in a relation
  of the form $\dot M \propto Z^{0.69}$ \citep{vin1:01}.}

For WR stars the situation is less clear because only a few models are
available. For late spectral subtypes \citet{vin1:05} recently claimed a
similar $Z$-dependence as for OB stars, with $\gamma = 0.86$.  As discussed
{\changedC previously}, our present WNL models show a qualitatively different
behaviour.  Apart from the influence of clumping, micro-turbulence, and high
wind density, it seems that models using a detailed CMF radiative transfer
give {\changedD results different from} {{\changedB wind models based on}} the Sobolev
approximation.  Without wind clumping both available {{\changedB CMF codes
    (CMFGEN by \citet{hil1:98} and our own PoWR code)}} tend to predict very
low wind accelerations $a_{\rm rad}$. The models can only account for the
observed wind momenta if clumping factors of the order of $D=10$ or larger are
assumed \citep{gra1:00,gra1:02,gra1:05,her1:01,hil1:03,hil2:03}. The reason
for this is not yet fully understood \citep[but see the discussion
in][]{hil2:03}.
In the following we try to quantify these differences by a detailed comparison
with previous works.

For comparison we use the $Z$-dependent mass loss prescription for OB stars by
\citet{vin1:01}, and the WNL models by \citet{vin1:05}.  Note that, although
the modeling of WNL subtypes was not particularly intended by
\citeauthor{vin1:01}, their model grid partly overlaps with our present
computations. In Fig.\,\ref{fig:zdep} we have extracted the $Z$-dependence of
our WNL models from Eqs.\,(\ref{eq:massloss1})--(\ref{eq:fit3}), for fixed
values of $\Gamma_{\rm e}$.
The blue curves indicate our results for Eddington parameters $\Gamma_{\rm
  e}$\,$=$\,$0.3$ to 0.9 (note that only the bold parts of the curves belong
to the parameter range where models have actually been computed). The stellar
parameters are fixed to $\log (T_\star/{\rm K})=4.65$,
$\log(L_\star/L_\odot)=6.3$, and $X_{\rm H}=0.4$. The red solid lines indicate
the mass loss predictions for OB stars by \citet{vin1:01} for the same stellar
parameters and the same range of Eddington factors.  The dashed red curve
represents the WNL models by \citet{vin1:05}, for slightly different
stellar parameters (see below).

The OB star models by \citet{vin1:01}
have a strictly exponential $Z$-dependence. Although $\Gamma_{\rm e}$ spans
the same parameter range as our WNL models, the resulting mass loss rates
cover a much narrower range, i.e., the dependence on $\Gamma_{\rm e}$ is much
weaker.
For large Eddington factors we obtain much higher mass loss rates, and a
flatter decline with $Z$. These important differences are most likely caused
by the different treatment of the wind dynamics in both codes.
\citeauthor{vin1:01} prescribe the velocity structure in their models and
solve for a global consistency of the wind energy.
The peculiar wind physics in the deep layers of WR winds, as described in
Sect.\,\ref{sec:thick}, is not considered in such an approach.

The WNL models by \citet{vin1:05} are indicated by the red dashed line in
Fig.\,\ref{fig:zdep}. They are computed for a similar effective temperature as
our models ($\log (T_\star/{\rm K})=4.60$), but for a lower hydrogen surface
abundance ($X_{\rm H}=0.15$), and a much lower luminosity
($\log(L_\star/L_\odot)=5.62$). The stellar mass is fixed to $20\,M_\odot$,
corresponding to $\Gamma_{\rm e}=0.37$, and the terminal wind speed is set to
$840\kms$.  For $Z < Z_\odot$ these models show exactly the same
$Z$-dependence as their OB star counterparts, while for $Z > Z_\odot$ the mass
loss tends to saturate.  Moreover, \citeauthor{vin1:05} detect a flattening of
the $Z$-dependence for $Z < 10^{-4} Z_\odot$, which is beyond the parameter
range investigated here. 

For $Z_\odot > Z > 10^{-4}Z_\odot$ {\changedC the WNL models by \citet{vin1:05}}
are in exact agreement with the mass loss prescription for OB stars by
\citet{vin1:01}. Note that \citet{vin1:01} assume a ratio of $\varv_\infty /
\varv_{\rm esc}$\,$=$\,2.6 for OB stars. With a typical stellar mass of
$43.5\,M_\odot$ for an OB star of this luminosity their mass loss prescription
gives $\dot M = 10^{-5.69} \msunpyr$ and $\varv_\infty = 2520\kms$ at solar
metallicity. A reduction of $\varv_\infty$ to $840\kms$ leads to an increase
of the mass loss rate by a factor of four, to $10^{-5.10} \msunpyr$.  A
reduction of the stellar mass to $20\,M_\odot$ together with an adjustment of
the Eddington factor to $\Gamma_{\rm e}=0.37$ further increases $\dot M$ by a
factor of 1.5, to $\dot M = 10^{-4.91} \msunpyr$. This is in exact agreement
with the WNL model by \citet{vin1:05} with $\dot M = 10^{-4.89}\msunpyr$.
Because the $Z$-dependence also is the same, the models by \citet{vin1:01}
thus closely resemble the WNL models by \citet{vin1:05}.

According to Eqs.\ (\ref{eq:massloss1})--(\ref{eq:fit3}) our own models give a
much lower value of $\dot M = 10^{-6.07}\msunpyr$.  Note, however, that our
formula is not reliably applicable because the desired Eddington factor of 0.37 is
very close to the parameter $\Gamma_0(Z_\odot) = 0.33$ {{\changedB (see the
    discussion at the end of Sect.\,\ref{sec:recipe})}}. A direct {{\changedB model
    computation}} yields $\dot M = 10^{-5.4}\msunpyr$ (see
Table\,\ref{tab:wrpar} for details).  {{\changedB The discrepancy is caused by the
    influence of the \HeII\ continuum, which effectively increases the
    Eddington factor close to the critical point in this specific parameter
    range}}.
Nevertheless, our mass loss rate is still significantly below the value from
\citeauthor{vin1:05}.

We conclude that our models are much more sensitive to the adopted
Eddington factor than others. This leads to higher or lower mass
loss rates, dependent on $\Gamma_{\rm e}$. According to our models the
proximity to the Eddington limit is the primary reason for the enhanced
WR-type mass loss. This possibility has also been discussed by
\citet{vin1:05}. Our models however show qualitative changes in the optically
thick regime that are also reflected in the $Z$-dependence of the mass loss.
While the other studies find a universal relation with $\dot{M} \propto
Z^{0.85}$ for almost all types of hot stars, we detect a flattening of the
$\dot M(Z)$ relation for $\Gamma_{\rm e} > 0.7$.  For lower values of
$\Gamma_{\rm e}$ our models show a similar $Z$-dependence {{\changedB to that determined
    by \citeauthor{vin1:05}}}. The important fact that our models yield high
mass loss rates over the whole range of metallicities implies that WR-type
mass loss might gain importance in metal-poor environments (see the discussion
in Sect.\,\ref{sec:lowz}).  }

\section{Discussion} 
\label{sec:discussion}

In the preceding sections we have presented the first fully self-consistent
models for the winds of WNL stars. These models reveal important information
about the nature of {\changedA these objects} and their mass loss.  In the
following we discuss the most important {\changedA points}: the reason for
their enhanced mass loss (Sect.\,\ref{sec:wrmassloss}), their evolutionary
state (Sect.\,\ref{sec:evolution}), the effect of wind clumping
(Sect.\,\ref{sec:dcl}), the role of radiative wind driving
(Sect.\,\ref{sec:inst}), the $Z$-dependence of WR-type mass loss
(Sect.\,\ref{sec:zdisc}), and its potential role in extremely metal-poor
environments (Sect.\,\ref{sec:lowz}).

\subsection{What is the reason for WR-type mass loss?}
\label{sec:wrmassloss}

The most important conclusion from our present work is that WR-type mass loss
is primarily triggered by the proximity to the Eddington limit.  {\changedA As
  the main reason we identify the extension of the deep atmospheric layers,
  leading to the formation of optically thick stellar winds. 
  We have performed the first detailed computations of the atmospheric
  structure of such objects, and we have identified qualitative differences to
  OB star winds. The most important is the strong dependence of the mass loss
  rates on the Eddington factor $\Gamma_{\rm e}$ and on the effective stellar
  temperature $T_\star$.  The idea that stellar mass loss increases close to
  the Eddington limit is of course not new. For example, for the case of LBVs
  \citet{vin1:02} also detected a strong dependence on $\Gamma_{\rm e}$.
  Nevertheless, in stellar evolution models the surface composition and
  temperature are commonly used to discriminate between different states of
  mass loss.}

{\changedA The proximity to the Eddington limit provides a natural explanation
  for the occurrence of the WR phenomenon, for He-burning stars with their
  inherently high $L/M$ ratios as well as for over-luminous H-burning stars.
  The latter is} expected for very massive stars at the end of their
main-sequence evolution and for fast rotators where fresh hydrogen is mixed
into the stellar core \citep{mey1:00}.  Note that such objects are also
expected to show WN surface compositions due to rotationally induced mixing
and mass loss.

The wind driving in our present models mainly originates from line opacities
of {\FeIII}--{\changedC\sc viii}. These ions belong to the ``cool'' Fe opacity peak, in
contrast to our previous WCE models, where the ``hot'' Fe peak (\FeIX--{\sc
  xvi}) is responsible for {\changedA the driving of the wind base}
\citep[][]{gra1:05}.  We thus confirm the prediction by \citet{nug1:02} that
the winds of WNL stars are driven by the cool Fe peak opacities, whereas for
early spectral subtypes the critical point is located in much deeper layers
where the hot Fe peak dominates.

\subsection{The evolutionary status of WNL stars}
\label{sec:evolution}

The dichotomy among the WN subtypes, where the H-rich WNL\,stars show much
higher luminosities than the earlier subtypes \citep[e.g.,][]{ham1:06},
indicates that the progenitors of {\changedC H-rich} WNL\,stars are very massive
stars with masses around $120\,M_\odot$.  In {\changedC Sect.\,\ref{sec:test}} we
have shown for the example of WR\,22 (WN\,7h) that the observed spectrum,
including the absolute flux distribution, is reproduced by a very luminous
model ($L_\star=10^{6.3}\,L_\odot$) with $\Gamma_{\rm e}=0.55$, corresponding
to a stellar mass of $78\,M_\odot$.  {\changedC Such a} high mass implies that
WR\,22 is still in the phase of central H-burning.  {\changedC Our models thus
  suggest that,} for very massive stars, the WNL phase might already occur at
the end of the main-sequence evolution, when the $L/M$ ratio increases due to the
increase of the mean molecular weight in the stellar core.  

Our parameters for WR\,22 are in agreement with the evolutionary model
for a rotating $120\,M_\odot$ star by \citet{mey1:03}, which is still in the
phase of central H-burning (at an age of $2.2\,10^6\,\mbox{yr}$). {\changedC This
  means that our derived mass suggests that WR\,22 is in the
  end phase of the main-sequence evolution. The observed mass loss rate of
  WR\,22 ($10^{-4.85}\msunpyr$) is not in agreement with the evolutionary
  models.  In the models the mass loss switches from O-star mass loss (in the
  range of $10^{-5.0}$--$10^{-4.65}\msunpyr$) to WNL mass loss
  ($\sim$\,$10^{-4}\msunpyr$) when the hydrogen surface mass fraction falls
  below 0.4. This sudden switch reduces the stellar mass very rapidly.  The
  previous evolution of WR\,22 has probably been different because, even with
  its present mass loss rate as the average value, it can only have lost up to
  $28\,M_\odot$ within 2\,Myr.}

{\changedC
  Our models suggest an H-rich WNL phase at the end of the main
  sequence evolution of very massive stars.  Within the standard evolutionary
  scenario such objects enter the LBV stage after leaving the main sequence
  \citep[e.g.,][]{cro1:07}. After losing their H-rich envelopes, they
  presumably evolve into the 'classical' He-burning WR stages.  Our models
  thus support an evolutionary sequence of the form O $\rightarrow$
  WNL\,(H-rich) $\rightarrow$ LBV $\rightarrow$ WN\,(H-poor) $\rightarrow$ WC
  for very massive stars, as proposed by \citet{lan1:94}.}
{\changedC According to our results in Sect.\,\ref{sec:test},} the mass loss from
WNL stars increases strongly {\changedC for} decreasing stellar temperatures
($\dot{M} \propto T_\star^{-3.5}$). For temperatures below the values examined
{\changedA here}, this might indeed lead to a smooth transition into the quiet
LBV phase.

{\changedC Additional evidence that the most massive stars are WNL stars comes from
  observations of young stellar clusters.}  There are {\changedC several cases
  where the brightest} stars in {\changedC such} clusters are WNL stars.  A good
example is the Carina OB cluster Tr\,16 where only two evolved stars are
present, the extremely luminous WNL star WR\,25 \citep[WN\,6h, see][]{ham1:06}
and $\eta$\,Car. Also the bright WNL stars in the Arches cluster
\citep{naj1:04,mar1:08} {\changedA and in R\,136a \citep{dek1:97}} fall into
the same category as WR\,22 and WR\,25 -- extremely luminous H-rich WNL stars
with relatively weak emission lines. {\changedC Moreover, the highest masses in
  binary systems are measured for WNL stars
  \citep[][]{rau1:96,sch1:99,bon1:04,mof1:07}. H-rich WNL stars thus} probably
mark an important phase of strong mass loss at the end of the main-sequence
evolution of very massive stars.

Note that we do not exclude the possibility of less luminous WNL stars being
in the phase of central He-burning, succeeding a RSG/LBV phase. Our present
assumption of high luminosities relies on galactic WNL stars with known
distance, which are usually members of dense clusters. There are, however,
many WNL stars that are not located in clusters, and that might have much
lower luminosities. Such a case is WR\,40 (WN\,8h) which is located in the
center of the ring nebula RCW\,58, a possible remnant of a previous RSG
outflow \citep[see, e.g.,][]{mar1:99}.

{\changedA
\subsection{The effect of wind clumping}
\label{sec:dcl}

Wind clumping is routinely taken into account in spectral analyses of WR
stars.  Its main effect is that the mean $\langle \rho^2 \rangle$ increases
and thus the derived mass loss rates, relying on $\rho^2$-diagnostics,
decrease with $\dot M \propto 1/\sqrt{D}$ \citep[see][]{ham1:98}. In the
present work we have shown that clumping also affects the wind acceleration.
The reason for this is that excited energy levels are {{\changedB more}}
populated by recombination processes. For the dense winds of WR stars, the
resulting increase of optically thin lines leads to an increase of the
radiative acceleration, approximately with $a_{\rm rad} \propto \sqrt{D}$.
Moreover, the force multiplier parameter $\alpha$ is reduced. At this point it
is necessary to emphasize that we assume small-scale clumps (micro-clumping)
in our models.  In this limit the separation between clumps is so small that
the mean opacity can be used in the radiative transfer. In the limit of large
clump separation (macro-clumping) the optical depth of individual clumps may
become large.  \citet{bro1:04} have shown that in this case the radiative
acceleration is reduced by geometrical effects. With a modified approach for
the wind opacity \citet{osk1:07} have shown that spectral diagnostics also are
considerably affected by macro-clumping.

We thus have two counteracting effects that may affect our results.
Micro-clumping increases the radiative acceleration with $a_{\rm rad} \propto
\sqrt{D}$, and macro-clumping decreases $a_{\rm rad}$ dependent on the
detailed clump-geometry. In our present models only micro-clumping is taken
into account. We have shown that it only affects the wind speed as long as it
is restricted to the outer wind region above the critical point. In some of
our models the critical region is however marginally affected by clumping.
This of course introduces an uncertainty in our mass loss rates.

Note that significant clumping factors in the range of $D$\,$=$\,10--100 have
recently been {{\changedB proposed for}} OB star winds
\citep[][]{cro2:02,hil2:03,bou1:03,bou1:05,ful1:06}.  If the effects of
micro-clumping dominate, the observed mass loss rates of OB stars thus should
be revised downward by factors of 3--10. This would of course introduce a
discrepancy in the previous wind models for OB stars because these do not take
clumping into account. As we discussed in Sect.\,\ref{sec:comp},
our models yield a much lower radiative force than others, which seems to be
connected to the detailed CMF radiative transfer. In contrast to the standard
Sobolev models, spectral lines have a finite width in our approach.
Interestingly, \citet{luc1:07} has proposed that the assumption of infinitely
narrow spectral lines leads to an over-estimation of the mass loss
rates in the standard models, i.e., OB star mass loss rates should actually be
lower. Because of its influence on the wind dynamics and the wind diagnostics,
it thus seems that clumping is fundamentally important for the physics of
radiatively driven winds.  Because the actual mechanisms of wind clumping are
presently not well understood, it introduces significant uncertainties in all
types of mass loss predictions for hot stars.}

\subsection{Are WR\,winds generally driven by radiation?}
\label{sec:inst}

In the present work we have shown that {\changedA WNL star winds can be
  explained self-consistently by pure radiative driving. As the reason for
  their strong optically thick winds we have identified the extension of the
  deep atmospheric layers due to the radiative force on Fe line opacities. For
  our present WNL models the cool Fe-peak opacities are responsible for this
  effect. Together with our previous results for early-type WC\,stars
  \citep{gra1:05}, where the inner wind is driven by the hot Fe-peak, one
  might now argue that WR winds are generally driven by radiation.}  This
would imply a general $Z$-dependence of WR mass loss.

{\changedA The fact that the two Fe opacity peaks act in different temperature
  regimes, however, has important consequences for our models. First, because
  of the gap between the two peaks, the mass loss cannot be maintained for
  intermediate stellar temperatures ($T_\star = 60$--100\,kK). Second, for
  cool objects, the hot Fe-peak disturbs the inner atmospheric structure.}

{\changedA The first point is connected to the well-known ``extension problem''
  of H-free WR stars. For H-free WN and WC stars stellar structure models
  predict much smaller radii than observed. Note that the predicted radii
  correspond to stellar temperatures above $\sim$\,100\,kK, in agreement with
  our wind models. Observed stellar temperatures, however, {{\changedB extend}}
  down to $\sim$\,50\,kK \citep[e.g.][]{ham1:06}. From a theoretical point of
  view such objects are thus not expected to exist, and they are not expected
  to have WR-type stellar winds.  \citet{ish1:99} have shown that very
  luminous WR stars might form an extended convection zone close to the
  stellar surface which can account for the observed extension. According to
  \citet{pet1:06} this extension however only occurs under the assumption of
  hydrostatic equilibrium. As soon as dynamic terms are included in their
  computations the extension disappears.  Interestingly, this extension is
  caused by the hot Fe-peak. According to our computations the hot Fe-peak
  drives the stellar wind, i.e., the possible extension of the outer layers of
  WR stars is hindered by their mass loss.  A possible solution for this
  problem could be the pulsational driving of the deep layers \citep[e.g., by
  strange mode instabilities as proposed by][]{gla1:02}. This could help to
  overcome the opacity-gap between the two Fe-peaks, and also account for the
  observed extension. Another possibility would be to bring the stars closer
  to the Eddington limit. For He-burning stars the $L/M$ ratio is however
  fixed. As also discussed by \citet{vin1:05}, fast stellar rotation may help
  to reduce the surface gravity of such objects and initiate strong mass loss.
  That this process works for optically thick winds, however, remains to be
  shown.  For OB star winds \citet{mad1:07} have demonstrated that rotation
  increases the mass loss at most by a factor of two.}

{\changedA The second problem only arises for late spectral subtypes. For these
  stars the cool Fe-peak drives the stellar wind, and the hot Fe-peak is
  located in deep layers at temperatures around 160\,kK.  For a star close to
  the Eddington limit it can thus happen that the radiative force exceeds
  gravity in the deep hydrostatic layers. In the stellar interior this would
  lead to the onset of convection.  Particularly for stars with dense winds
  this situation already occurs at the inner boundary of our
  atmosphere models.  We thus reach a mass loss limit above which we cannot
  obtain stationary wind models. In reality this presumably leads} to a
non-stationary situation.  \citet{dor1:06} have recently shown that the
observed variability in WN\,8 subtypes can indeed be explained by
instabilities that are caused by the hot Fe-peak.

The comparison with observations in Sect.\,\ref{sec:test} has shown that only
a part of the galactic WR sample is reproduced by our models.  {\changedA
  The WN\,8 subtypes particularly have higher wind densities than predicted.
  The comparison with the WN\,8 model from \citet{vin1:05} in
  Sect.\,\ref{sec:comp} has shown that this problem still holds if lower
  luminosities are adopted. According to our models a different driving
  mechanism thus would be desirable, at least for the wind base of these
  objects. Again, this mechanism could be related to the observed pulsations
  \citep[see e.g.][]{mar1:98,lef1:05} or rotational driving.  Note that also
  the non-detection of X-ray emission from WN\,8 subtypes might be a hint to a
  different wind driving mechanism \citep[see][]{osk1:05}.}

{\changedA
\subsection{The $Z$-dependence of WR mass loss}
\label{sec:zdisc}

As expected for radiatively driven winds, our models show a strong
$Z$-dependence. Because of the important influence of the Eddington factor, it
is however not sufficient to parametrize the mass loss as a function of $Z$
alone. In Sect.\,\ref{sec:recipe} we have thus prepared a mass loss recipe
which takes the combined effect of $\Gamma_{\rm e}$ and $Z$ into account.
According to Eqs.\,(\ref{eq:massloss})--(\ref{eq:fit3}) the metallicity $Z$
affects the resulting mass loss rates in two ways. First, the parameter
$\Gamma_0(Z)$ decreases for increasing $Z$ (Eq.\,\ref{eq:fit3}).  $\Gamma_0$
characterizes the limiting Eddington factor for which WR-type mass loss sets
in.  An increasing metallicity thus brings the star effectively closer to the
Eddington limit, and supports the formation of WR-type winds.  Second, the
exponent $\beta(Z)$ increases with $Z$ (Eq.\,\ref{eq:fit2}). This results in
the flattening of the $\dot M(Z)$ relation for low $Z$ and large
$\Gamma_{\rm e}$ (see Sect.\,\ref{sec:comp}).

A higher metallicity thus has two effects. The mass loss from WR stars
increases and WR stars are formed more easily. According to our models, at
solar metallicity the most massive stars already enter the WNL phase at the
end of the H-burning stage.  For $Z = 3\,Z_\odot$ we could even show that
stars with $120\,M_\odot$ already become WNL stars on the {\changedD zero-age
  main sequence} \citep{gra5:06}. In low metallicity environments, on the other
  hand, the occurrence of WR-type winds should be restricted to He-burning
  objects and maybe to fast rotators.  The influence of metallicity on WR
  populations is thus extremely complex. In different environments different
  objects become WR stars.}

{\changedA
\subsection{WR-type mass loss in extremely metal-poor environments}
\label{sec:lowz}

In Sect.\,\ref{sec:comp} we have shown that the slope of the $\dot M(Z)$
relation becomes {\changedD shallower} for increasing $\Gamma_{\rm e}$. Close to the
Eddington limit our models are thus able to produce large mass loss rates
also at very low $Z$. Note that these are presently the only wind models for
low metallicities for which such an efficient mass loss mechanism has been
detected.
The large Eddington factors that are required ($\Gamma_{\rm e} \approx 0.8$)
are however only reached by very massive objects or stars close to critical
rotation.  Both types of objects are indeed expected at low $Z$ \citep[see
also the discussion in][]{vin1:05}. For extremely low metallicities ($Z <
10^{-4}\,Z_\odot$) more massive stars are expected to form 
\citep{bro1:99,nak1:02}. Moreover, evolutionary calculations indicate that
rotating massive stars tend to approach critical rotation more easily at low
$Z$ \citep{mey1:02}. Our present models show that high mass loss rates can be
maintained in this regime, and that such objects should appear as WR\,stars
with particularly low wind velocities. Note, however, that the question of whether
rotation really leads to increased mass loss remains to be investigated in
more detail. For CAK-type winds \citet{mad1:07} recently demonstrated that
rotation increases $\dot M$ at most by a factor of two.}

A new and very important result from our calculations concerns the role of
primary elements for the wind driving of extremely metal-poor stars.
{\changedA \citet{yoo1:05}, as well as \citet{mey1:06}, have shown that
  rotating massive stars at extremely low $Z$\ ($10^{-5}\,Z_\odot$ and
  $10^{-8}\,Z_\odot$ respectively) are capable of producing nitrogen mass
  fractions in the range of one percent at their surface.
  The mass loss from such objects may potentially enrich the early ISM with
  freshly produced nitrogen. 
With the adopted stellar yields from \citet{mey1:06}, 
\citet{chi1:06} could explain  the observed high
nitrogen abundances in extremely metal-poor halo stars
\citep{spi1:05}.}

Our iron-free models in Sect.\,\ref{sec:zdep} now confirm that such
self-enriched objects can develop strong stellar winds. We find that iron-free
{\changedA objects} with solar $Z$ (i.e., $X_{\rm CNO}=Z_\odot$ and $X_{\rm
  Fe}=0$) {\changedA can reach} mass loss rates similar to stars with standard
composition at $Z_\odot/50$.  This means that medium-complex elements like C,
N, and O are much less efficient for the wind driving than the complex
Fe-group elements.  For more realistic mass loss predictions it is thus
important to take the relative contributions of different elements into
account.

For the application of $Z$-dependent mass loss relations like our
Eqs.\,(\ref{eq:massloss}), 
(\ref{eq:fit2}), and (\ref{eq:fit3}), we thus suggest the use of an effective
metallicity $Z_{\rm eff}$, where the relative contributions of different
elements are adequately weighted. In the present case of CNO-enriched WNL
stars we propose an {{\changedB approximation}} of the form
\begin{equation}
Z_{\rm eff} \approx Z_\odot \cdot 
\left( \frac{1}{50} \frac{X_{\rm CNO}}{X_{\rm CNO, \odot}}
      +\frac{X_{\rm Fe}}{X_{\rm Fe, \odot}} \right).
\end{equation}

\section{Conclusions} 
\label{sec:conclusions}

We have performed a systematic study of the mass loss from late-type WN stars
by means of a new generation of non-LTE atmosphere models, including a
self-consistent treatment of the wind hydrodynamics. This was the first
application of these complex models on a large scale. We could show that the
strong winds from WNL stars can be explained {\changedA self-consistently} by
radiative driving, and that the proximity to the Eddington limit is the
primary trigger of the enhanced WR-type mass loss. {\changedA Moreover we have
  identified important qualitative differences to OB star winds that are
  caused by the altered wind physics.}

{{\changedB Our models}} reproduce the observed sequence of late spectral
subtypes from WN\,6 to WN\,9 {\changedC qualitatively}.  {{\changedB The obtained
    wind densities are in agreement with the galactic WNL sample, except for
    the enigmatic WN\,8 subtypes with strong winds. A detailed}} comparison
with the spectroscopic binary WR\,22
{\changedC supports} our key statement that H-rich WNL stars are very {{\changedB
    massive, luminous stars}} close to the Eddington limit.  {{\changedB Our
    models imply that these objects are still in the phase}} of central
H-burning, {\changedC in line with an evolutionary sequence of the form\, O
  $\rightarrow$ WNL\,(H-rich) $\rightarrow$ LBV $\rightarrow$ WN\,(H-poor)
  $\rightarrow$ WC\, for very massive stars \citep[e.g.,][]{lan1:94,cro1:07}.}
Due to the strong temperature-dependence of the mass loss ($\dot{M} \propto
T_\star^{-3.5}$), an evolution towards cooler effective temperatures may
indeed lead to a continuous transition into the quiet LBV phase.

As expected for radiatively driven winds, our models show a strong
$Z$-dependence. The influence of the Eddington factor $\Gamma_{\rm e}$
{\changedA however} is at least equally important. We find that
the $\dot{M}(Z)$ relation for WNL\,stars becomes flatter for higher
$\Gamma_{\rm e}$, i.e., high mass loss rates can also be {\changedC maintained for}
very low $Z$. Our models thus predict an efficient mass loss mechanism for low
metallicity stars. For stars in extremely metal-poor environments (i.e.\ the
second generation of massive stars) we find that the surface-enrichment with
primary nitrogen, as {\changedB described, e.g.}\ by \citet[][]{mey1:06}, may
lead to strong mass loss.  We thus confirm that these first WN\,stars might
play a key role in the enrichment of the early ISM with freshly produced
nitrogen.


\end{document}